\documentclass[twocolumn]{aastex701}

\usepackage{CJKutf8}

\received{Jan 14, 2026}

\submitjournal{AJL}

\shorttitle{The XRD}
\shortauthors{Hviding et al.}

\graphicspath{{figures}}

\begin{document}

\title{The X-Ray Dot: Exotic Dust or a Late-Stage Little Red Dot?}

\author[orcid=0000-0002-4684-9005]{Raphael E. Hviding}
\affiliation{Max-Planck-Institut f\"ur Astronomie, K\"onigstuhl 17, D-69117 Heidelberg, Germany}
\email[show]{hviding@mpia.de}

\author[orcid=0000-0002-2380-9801]{Anna de Graaff}\thanks{Clay Fellow}
\affiliation{Center for Astrophysics, Harvard \& Smithsonian, 60 Garden St, Cambridge, MA 02138, USA}
\affiliation{Max-Planck-Institut f\"ur Astronomie, K\"onigstuhl 17, D-69117 Heidelberg, Germany}
\email{}

\author[orcid=0000-0003-2488-4667]{Hanpu Liu (\begin{CJK}{UTF8}{gbsn}刘翰溥\end{CJK})}
\affiliation{Department of Astrophysical Sciences, Princeton University, Princeton, NJ 08544, USA}
\email{hanpu.liu@princeton.edu}

\author[orcid=0000-0003-4700-663X]{Andy D. Goulding}
\affiliation{Department of Astrophysical Sciences, Princeton University, Princeton, NJ 08544, USA}
\email{goulding@astro.princeton.edu}

\author[orcid=0000-0002-0463-9528]{Yilun Ma (\begin{CJK}{UTF8}{gbsn}马逸伦\end{CJK})}
\affiliation{Department of Astrophysical Sciences, Princeton University, Princeton, NJ 08544, USA}
\email{yilun@princeton.edu}

\author[orcid=0000-0002-5612-3427]{Jenny E. Greene}
\affiliation{Department of Astrophysical Sciences, Princeton University, Princeton, NJ 08544, USA}
\email{jgreene@astro.princeton.edu}

\author[0000-0002-3952-8588]{Leindert A. Boogaard} \affiliation{Leiden Observatory, Leiden University, PO Box 9513, NL-2300 RA Leiden, The Netherlands}
\email{boogaard@strw.leidenuniv.nl}

\author[0000-0002-8651-9879]{Andrew J. Bunker} \affiliation{Department of Physics, University of Oxford, Denys Wilkinson Building, Keble Road, Oxford OX1 3RH, UK}
\email{andy.bunker@physics.ox.ac.uk}

\author[0000-0001-7151-009X]{Nikko J. Cleri}
\affiliation{Department of Astronomy \& Astrophysics, The Pennsylvania State University, University Park, PA 16802, USA}
\affiliation{Institute for Computational \& Data Sciences, The Pennsylvania State University, University Park, PA 16802, USA}
\affiliation{Institute for Gravitation and the Cosmos, The Pennsylvania State University, University Park, PA 16802, USA}
\email{cleri@psu.edu}

\author[0000-0002-8871-3026]{Marijn Franx} \affiliation{Leiden Observatory, Leiden University, PO Box 9513, NL-2300 RA Leiden, The Netherlands}
\email{franx@strw.leidenuniv.nl}

\author[0000-0002-3301-3321]{Michaela Hirschmann}
\affiliation{Institute for Physics, Laboratory for Galaxy Evolution and Spectral modelling, Ecole Polytechnique Federale de Lausanne, Observatoire de Sauverny, Chemin Pegasi 51, 1290 Versoix, Switzerland}
\email{michaela.hirschmann@epfl.ch}

\author[0000-0001-6755-1315]{Joel Leja}
\affiliation{Department of Astronomy \& Astrophysics, The Pennsylvania State University, University Park, PA 16802, USA}
\affiliation{Institute for Computational \& Data Sciences, The Pennsylvania State University, University Park, PA 16802, USA}
\affiliation{Institute for Gravitation and the Cosmos, The Pennsylvania State University, University Park, PA 16802, USA}
\email{joel.leja@psu.edu}

\author[0000-0003-3729-1684]{Rohan P. Naidu}\thanks{MIT Pappalardo Fellow}
\affiliation{MIT Kavli Institute for Astrophysics and Space Research, 70 Vassar Street, Cambridge, MA 02139, USA}
\email{rnaidu@mit.edu}

\author[0000-0003-2871-127X]{Jorryt Matthee} \affiliation{Institute of Science and Technology Austria (ISTA), Am Campus 1, 3400 Klosterneuburg, Austria}
\email{jorryt.matthee@ist.ac.at}

\author[orcid=0000-0003-4075-7393]{David J. Setton}\thanks{Brinson Prize Fellow}
\affiliation{Department of Astrophysical Sciences, Princeton University, Princeton, NJ 08544, USA}
\email{davidsetton@princeton.edu}

\author[orcid=0000-0003-4891-0794]{Hannah Übler}
\affiliation{Max-Planck-Institut für extraterrestrische Physik, Gießenbachstraße}
\email{hannah@mpe.mpg.de}

\author[0000-0001-8349-3055]{Giacomo Venturi}
\affiliation{Scuola Normale Superiore, Piazza dei Cavalieri 7, I-56126 Pisa, Italy}
\email{giacomo.venturi1@sns.it}

\author[0000-0001-9269-5046]{Bingjie Wang (\begin{CJK}{UTF8}{gbsn}王冰洁\end{CJK})}\thanks{NHFP Hubble Fellow} 
\affiliation{Department of Astrophysical Sciences, Princeton University, Princeton, NJ 08544, USA}
\email{bjwang@princeton.edu}

\begin{abstract}

JWST's ``Little Red Dots'' (LRDs) are increasingly interpreted as active galactic nuclei (AGN) obscured by dense thermalized gas rather than dust as evidenced by their X-ray weakness, blackbody-like continua, and Balmer line profiles.
A key question is how LRDs connect to standard UV-luminous AGN and whether transitional phases exist and if they are observable.
We present the ``X-Ray Dot'' (XRD), a compact source at $z=3.28$ observed by the NIRSpec WIDE GTO survey.
The XRD exhibits LRD hallmarks: a blackbody-like ($T_{\rm eff} \simeq 6400$\,K) red continuum, a faint but blue rest-UV excess, falling mid-IR emission, and broad Balmer lines ($\rm FWHM \sim 2700-3200\,km\,s^{-1}$).
Unlike LRDs, however, it is remarkably X-ray luminous ($L_\textrm{2$-$10\,keV} = 10^{44.18}$\,erg\,s$^{-1}$) and has a continuum inflection that is bluewards of the Balmer limit.
We find that the red rest-optical and blue mid-IR continuum cannot be reproduced by standard dust-attenuated AGN models without invoking extremely steep extinction curves, nor can the weak mid-IR emission be reconciled with well-established X-ray--torus scaling relations.
We therefore consider an alternative scenario: the XRD may be an LRD in transition, where the gas envelope dominates the optical continuum but optically thin sightlines allow X-rays to escape.
The XRD may thus provide a physical link between LRDs and standard AGN, offering direct evidence that LRDs are powered by supermassive black holes and providing insight into their accretion properties.

\end{abstract}


\section{Introduction}\label{sec:intro}

JWST's ``Little Red Dots'' (LRDs; \citealt{matthee_LittleRedDots_2024}) are a ubiquitous population of high-redshift ($z \gtrsim 4$) compact sources with unique spectro-photometric properties \citep[e.g.,][]{kokorev_CensusPhotometricallySelected_2024,kocevski_RiseFaintRed_2025,hviding_RUBIESSpectroscopicCensus_2025}.
Characterized by V-shaped rest--UV-optical spectral energy distributions (SEDs) and broad Balmer emission lines, these objects were, at least in part, initially interpreted as dust-reddened Active Galactic Nuclei (AGN) \citep[e.g.,][]{matthee_LittleRedDots_2024}.
However, LRDs are notably X-ray undetected, remaining faint even in stacks \citep[e.g.,][]{yue_StackingXRayObservations_2024,ananna_XRayViewLittle_2024}, and lack the strong mid-infrared detections expected from a hot dust torus \citep[e.g.,][]{williams_GalaxiesMissedHubble_2024,casey_DustLittleRed_2024,akins_COSMOSWebOverabundancePhysical_2025,setton_ConfirmedDeficitHot_2025}.
Furthermore, many LRDs exhibit strong spectral breaks near the Balmer limit, extreme Balmer decrements ($>10$), and high rates of Balmer line absorbers \citep[$>30\%$;][]{matthee_LittleRedDots_2024,setton_LittleRedDots_2024,juodzbalis_JADESRosettaStone_2024,wang_RUBIESEvolvedStellar_2024,deugenio_BlackTHUNDERStrikesTwice_2025,degraaff_LittleRedDots_2025,torralba_WarmOuterLayer_2025} that are difficult to reconcile with dust attenuation models.

A new paradigm proposes that LRDs are AGN embedded in dense, optically thick gas rather than dust. 
In this ``black hole star'' scenario, the red continuum originates from an optically thick gas envelope with similarities to a stellar atmosphere, yielding a Balmer break via absorption from collisionally excited hydrogen populating the $n=2$ level \citep{inayoshi_ExtremelyDenseGas_2025,naidu_BlackHoleStar_2025,degraaff_RemarkableRubyAbsorption_2025,ji_BlackTHUNDERNonstellarBalmer_2025}.
Moreover, additional dense-gas signatures are seen in Balmer line profiles, including line-of-sight absorbers \citep[e.g.,][]{juodzbalis_JADESRosettaStone_2024,deugenio_BlackTHUNDERStrikesTwice_2025} and extended wings that are sometimes consistent with electron-scattering broadening \citep[e.g.,][]{rusakov_JWSTsLittleRed_2025,chang_ImpactResonanceRaman_2025}. 
These features need not arise from a single component, but instead can reflect a multiphase medium with distinct emitting, absorbing, and scattering regions.
Such gas-rich environments can naturally suppress X-ray emission via high column densities, although super-Eddington accretion has also been proposed to explain the X-ray weakness of LRDs through intrinsically steeper X-ray spectra \citep[e.g.,][]{tortosa_SystematicBroadbandXray_2023,madau_XRayWeakActive_2024,lambrides_CaseSupereddingtonAccretion_2024}.

If LRDs indeed represent a phase of rapid, gas-enshrouded accretion, they may play a central role in the mass assembly of supermassive black holes (SMBHs) and potentially reconcile the discovery of massive SMBHs in the early Universe \citep[$z\gtrsim5$; e.g.,][]{inayoshi_LittleRedDots_2025}.
However, a key open question remains: how are LRDs linked to the well-studied population of X-ray and UV-luminous accreting SMBHs at later cosmic times?
While stacking analyses indicate that a large fraction of LRDs are X-ray weak, it remains unclear whether this applies to the entire population, particularly given their relative faintness.
Alternatively, the sharp decline in the number density of LRDs around $z\sim3$ \citep{kokorev_CensusPhotometricallySelected_2024,kocevski_RiseFaintRed_2025,ma_CountingLittleRed_2025} may suggest a transition in accretion mode or obscuration, raising the question of whether an intermediate evolutionary phase should be observable as LRDs evolve into the UV-luminous quasar population.

\begin{figure*}[ht!]
    \centering
    \includegraphics[width=\textwidth]{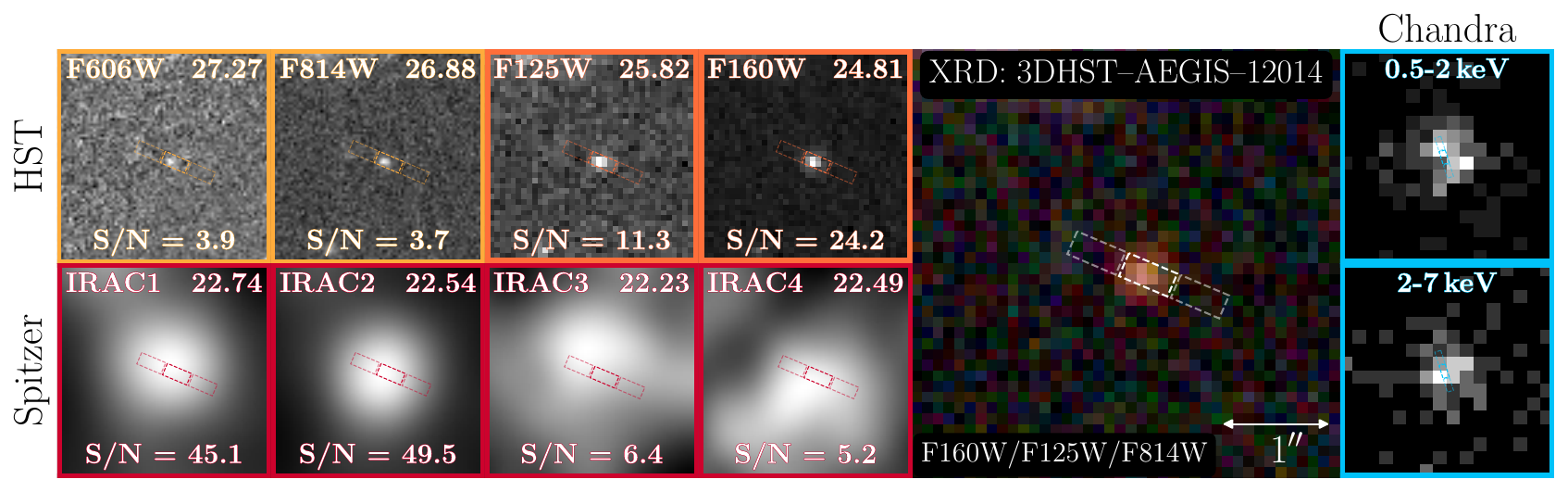}
    \includegraphics[width=\textwidth]{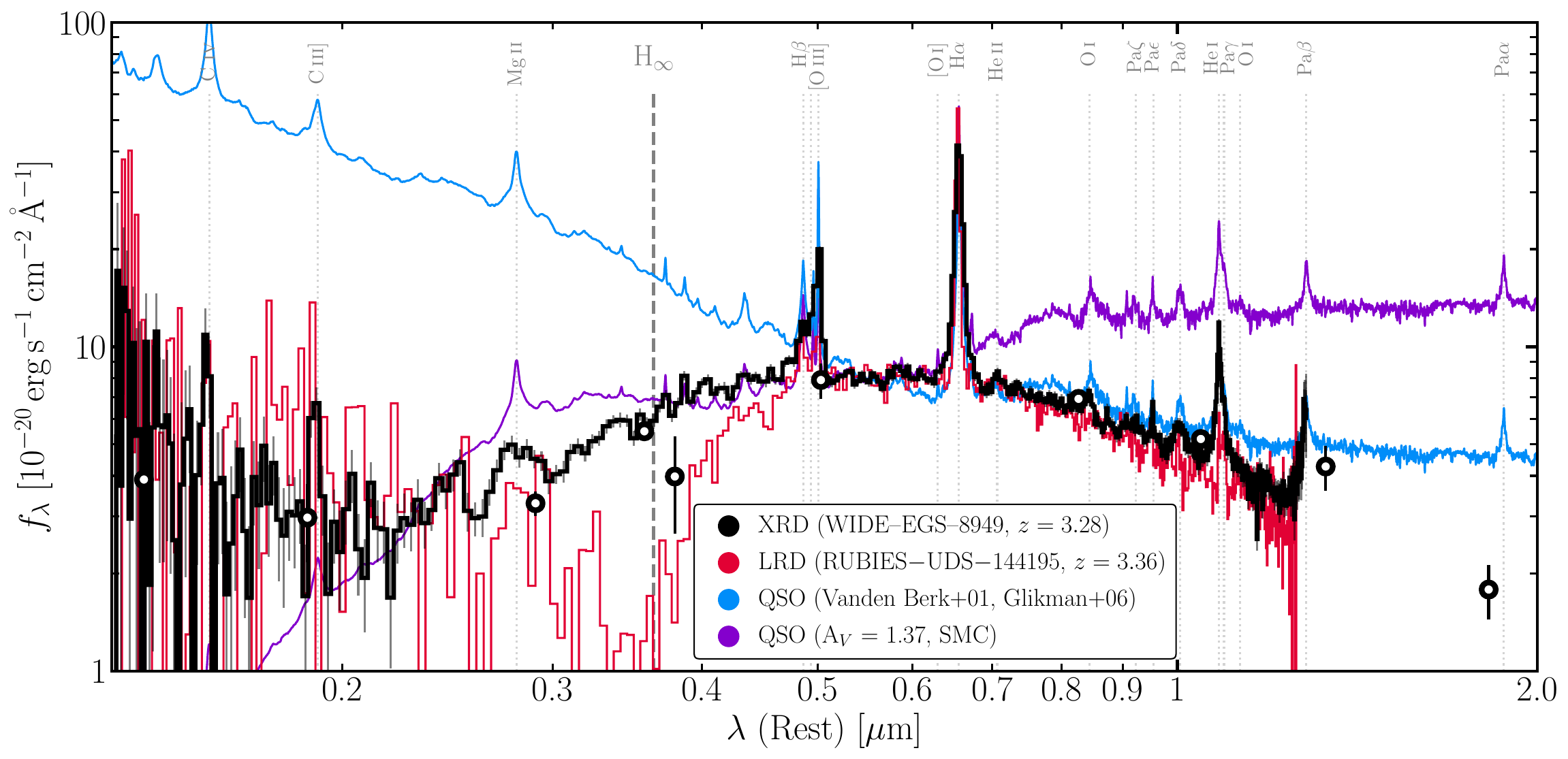}
    \caption{
    Overview of observations for the X-Ray Dot (XRD). 
    Top panels present the space-based imaging compiled from the 3DHST survey from HST and Spitzer, the RGB composite from HST, and the hard and soft X-ray cutouts from Chandra.
    HST/Spitzer panels show the $4''\times4''$ cutout in the band, magnitude, and S/N whereas the Chandra panels show $8''\times8''$ cutouts.
    The bottom panel compares the XRD photometry and scaled spectrum (black) to an LRD with a similar rest-optical continuum (RUBIES$-$UDS$-$144195, red) and to the STScI composite quasar spectrum derived from \citet{vandenberk_CompositeQuasarSpectra_2001} and \citet{glikman_NearInfraredSpectralTemplate_2006} shown unattenuated (blue) and extincted (purple; $\rm A_{V}=1.37$) by an SMC-like dust law \citep{gordon_QuantitativeComparisonSmall_2003}.
    Comparison spectra are normalized to the XRD scaled spectrum at rest 5500\AA.
    }
    \label{fig:xrd}
\end{figure*}

In the black hole star scenario, the envelope should eventually disperse as the SMBH accretion rate exceeds that of the gas infall \citep{kido_BlackHoleEnvelopes_2025}.
One may therefore expect a transitionary phase where the X-rays can escape as sightlines to the accreting black hole begin to clear.
Recently, \citet{fu_DiscoveryTwoLittle_2025} identified two LRD candidates at $z\sim2.9$, dubbed the ``Forges'', that exhibit broad rest near-IR emission lines and a photometric V-shape alongside intense X-ray, radio, and mid-IR emission which they suggest mark a transitionary LRD phase with elements of both UV-luminous quasars and LRDs.
However, unlike any LRD observed to date, the sources reported by \citet{fu_DiscoveryTwoLittle_2025} show strong evidence for hot dust emission. 
Moreover, their bolometric luminosities ($\sim10^{46}$\,erg\,s$^{-1}$) are $>1$ dex higher than typical LRDs ($\sim10^{43-45}$\,erg\,s$^{-1}$), raising the question of where precisely these objects fit into LRD evolution. 

In this Letter, we present 3DHST-AEGIS-12014, hereafter the ``X-Ray Dot'' (XRD), a compact source at $z=3.28$ with a blackbody-like red rest-optical continuum ($L_{\rm bol}\sim 5\times10^{44}$\,erg\,s$^{-1}$), broad Balmer emission lines, and a falling mid-IR continuum, all features typically associated with LRDs. 
Notably, the XRD exhibits luminous, and potentially variable, X-ray emission.
In Section~\ref{sec:xrd}, we present the existing CFHT/HST/Spitzer photometric data, NIRSpec/PRISM spectroscopy obtained as a part of the GTO WIDE survey \citep{maseda_NIRSpecWideGTO_2024}, and Chandra X-ray observations used in this work. 
We first consider the scenario that this source is a dust-reddened AGN in Section~\ref{sec:dust}, but show that standard dust prescriptions fail to explain the XRD's properties.
We therefore explore the possibility that the XRD is a late-stage LRD, and discuss the implied physical conditions of the gas in Section~\ref{sec:gas}.
Finally, we present our summary and discussion in Section~\ref{sec:discussion}.
We assume a flat $\Lambda$CDM cosmology with $\Omega_{\rm m}=0.3$ and $H_0=70$\,km\,s$^{-1}$\,Mpc$^{-1}$. 

\section{The X-Ray Dot}\label{sec:xrd}

\subsection{Photometry}\label{ssec:phot}

The All-Wavelength Extended Groth Strip International Survey \citep[AEGIS;][]{davis_AllWavelengthExtendedGroth_2007} provides a wealth of photometry from the X-ray through sub-mm across the Extended Groth Strip (EGS). 
We obtain optical through mid-IR photometry of the XRD spanning $0.6-24\,\micron$ (CFHT WIRCam, HST ACS and WFC3, and Spitzer IRAC and MIPS) from the public 3D-HST photometric catalogs \citep{brammer_3DHSTWidefieldGrism_2012,skelton_3DHSTWFC3selectedPhotometric_2014}, plotted in Figure~\ref{fig:xrd} and presented in Table~\ref{tab:xrd}.
In addition, we find no sub-mm detection in the SCUBA-2 Cosmology Legacy Survey and assume a flux of zero with an error of 1.2\,mJy based on the average $1\sigma$ depth in the EGS \citep{geach_SCUBA2CosmologyLegacy_2017}.
We note that upcoming NOEMA observations (PI: Hviding) of the XRD will provide deeper sub-millimeter constraints, further placing constraints on the cold dust energy budget.

The source is extremely compact, with an effective radius of $r_e = 0.0322\pm0.0042''$ in F160W from the \citet{vanderwel_StructuralParametersGalaxies_2012} \texttt{GALFIT} catalog, where it is flagged as a good fit.
Given the F160W PSF FWHM of $\approx 0.18''$, this size is consistent with being unresolved and corresponds to a physical size of $r_e \lesssim 250$\,pc.
We note that the source appears marginally resolved in the rest-UV ACS imaging (Figure~\ref{fig:xrd}), although a detailed morphological analysis would require high-SNR rest-UV NIRCam imaging. 
Extended rest-UV emission may be consistent with similar emission seen in other LRDs that may arise from a host galaxy \citep[e.g.,][]{rinaldi_NotJustDot_2024,torralba_WarmOuterLayer_2025}.

\begin{table}[ht!]
\caption{X-Ray Dot Measurements}
\begin{center}
\begin{tabular}{lcc}
\hline
\hline
\hline
\multicolumn{3}{c}{Position} \\
\multicolumn{3}{c}{$z = 3.278$} \\
\multicolumn{3}{c}{$\rm14^h20^m47.498^s$ $\rm+53^\circ02^\prime32.83^{\prime\prime}$} \\
\hline
\hline
\multicolumn{3}{c}{Photometry} \\
\multicolumn{1}{c}{Band} & Flux [$\mu$Jy] & Error [$\mu$Jy] \\
\hline
HST ACS F606W & 0.044 & 0.011 \\
HST ACS F814W & 0.064 & 0.017 \\
\hline
HST WFC3 F125W & 0.169 & 0.015 \\
HST WFC3 F160W & 0.430 & 0.018 \\
\hline
CFHT WIRCam J & $-$0.043 & 0.093 \\
CFHT WIRCam H & 0.35 & 0.12 \\
CFHT WIRCam Ks & 1.22 & 0.15 \\
\hline
Spitzer IRAC Ch.~1 & 2.885 & 0.064 \\
Spitzer IRAC Ch.~2 & 3.477 & 0.070 \\
Spitzer IRAC Ch.~3 & 4.62 & 0.72 \\
Spitzer IRAC Ch.~4 & 3.62 & 0.70 \\
\hline
Spitzer MIPS 24$\mu$m & 4.4 & 5.1 \\
\hline
\hline
\multicolumn{3}{c}{X-Ray Analysis} \\
\multicolumn{1}{c}{Parameter} & Measurement & Error \\
\hline
Exposure Time [s] & $7.729\times10^5$ & --- \\
Aperture Counts [ct] & 205.0 & 14.2 \\
Background Counts [ct] & 15.818 & 0.460 \\
Net Counts [ct] & 189.182 & 14.325 \\
Rate (0.5$-$7\,keV) [ct\,s$^{-1}$] & $2.46\times10^{-4}$  & $0.19\times10^{-4}$ \\
$f_\textrm{0.5$-$7\,keV}$ [erg\,s$^{-1}$\,cm$^{-2}$] & $2.7\times10^{-15}$ & $0.2\times10^{-15}$ \\
Power-law Index ($\Gamma$) & 1.8 & 0.2 \\
$N_\textrm{H}$ (XRD) [cm$^{-2}$] & $2.4\times10^{22}$ & $2.0\times10^{22}$ \\
$L_\textrm{2$-$10\,keV}$ (Int.) [erg\,s$^{-1}$] & $1.51\times10^{44}$ & $0.24\times10^{44}$ \\
Cash Statistic ($\mathrm{DoF}=441$)&  321.3 & --- \\
\hline
\hline
\hline
\end{tabular}
\label{tab:xrd}
\end{center}
\end{table}

\subsection{JWST NIRSpec Spectroscopy}\label{ssec:nirspec}

The NIRSpec Wide GTO Survey \citep{maseda_NIRSpecWideGTO_2024} is a 100-hour Cycle 1 program providing shallow low- ($\mathcal{R}\sim100$) and high-resolution ($\mathcal{R}\sim2700$) NIRSpec/MSA spectroscopy across all five CANDELS fields, with a target selection primarily based on the 3DHST catalogs. 
The XRD (3DHST-AEGIS-12014, WIDE-EGS-8949) was targeted with the highest priority (Priority Class 1) because of its high X-ray luminosity. 
The obtained PRISM, G235H, and G395H spectra have total exposure times of 40\,min, 27\,min, and 29\,min respectively.
We use the reduced NIRSpec MSA spectra from version 4.4 of the public DAWN JWST Archive \citep[DJA;][]{brammer_DAWNJWSTArchive_2025}, processed with \texttt{msaexp} \citep{brammer_MsaexpNIRSpecAnalyis_2023} following the methodology described in \citet{heintz_JWSTPRIMALArchivalSurvey_2025} and \citet{degraaff_RUBIESCompleteCensus_2025}.
The 1D spectra were extracted using optimal extraction \citep{horne_OptimalExtractionAlgorithm_1986}, with a kernel used to correct for wavelength-dependent slit losses based on the source position.

We show the PRISM spectrum of the XRD in Figure~\ref{fig:xrd} alongside a typical LRD with a strong Balmer break at similar redshift from the RUBIES survey (RUBIES-UDS-144195; \citealt{degraaff_RUBIESCompleteCensus_2025}). 
We also show the STScI composite quasar template derived from \citet{vandenberk_CompositeQuasarSpectra_2001} and \citet{glikman_NearInfraredSpectralTemplate_2006} as both unobscured and obscured by an SMC dust law \citep{gordon_QuantitativeComparisonSmall_2003}. 
Although the XRD shares similarities with both populations, such as the red rest-optical continuum of LRDs, it also exhibits distinct differences in its spectral shape and mid-IR properties which we discuss in detail in Section~\ref{sec:dust}.
Comparing the spectrum to the photometry, we find that the PRISM spectrum is a factor of 0.75 fainter.
This discrepancy may arise from a combination of factors: (1) the HST-based astrometry may limit the pointing accuracy and hence the applied slit loss correction; (2) the slit loss correction itself may carry systematic uncertainties; and (3) given the source has weak evidence for X-ray variability (Appendix \ref{app:xray}), we may expect some variability in the rest-UV/optical over the $\sim$decade time baseline between the HST and JWST observations.
We apply this correction throughout this work where noted, but emphasize that NIRCam observations of this source will be critical to robustly anchor the fluxes.

We perform emission line fitting using \texttt{unite} \citep{hviding_TheSkyentistUniteVersion_2025}, which fits the PRISM, G235H, and G395H spectra simultaneously. 
This approach accounts for the wavelength dependence and undersampling of the line-spread function across the different dispersers. 
We model all emission lines with a narrow component sharing a common width.
For permitted transitions, we test additional broad components using Gaussian, Lorentzian, and exponential profiles. 

\begin{figure}
    \centering
    \includegraphics[width=\columnwidth]{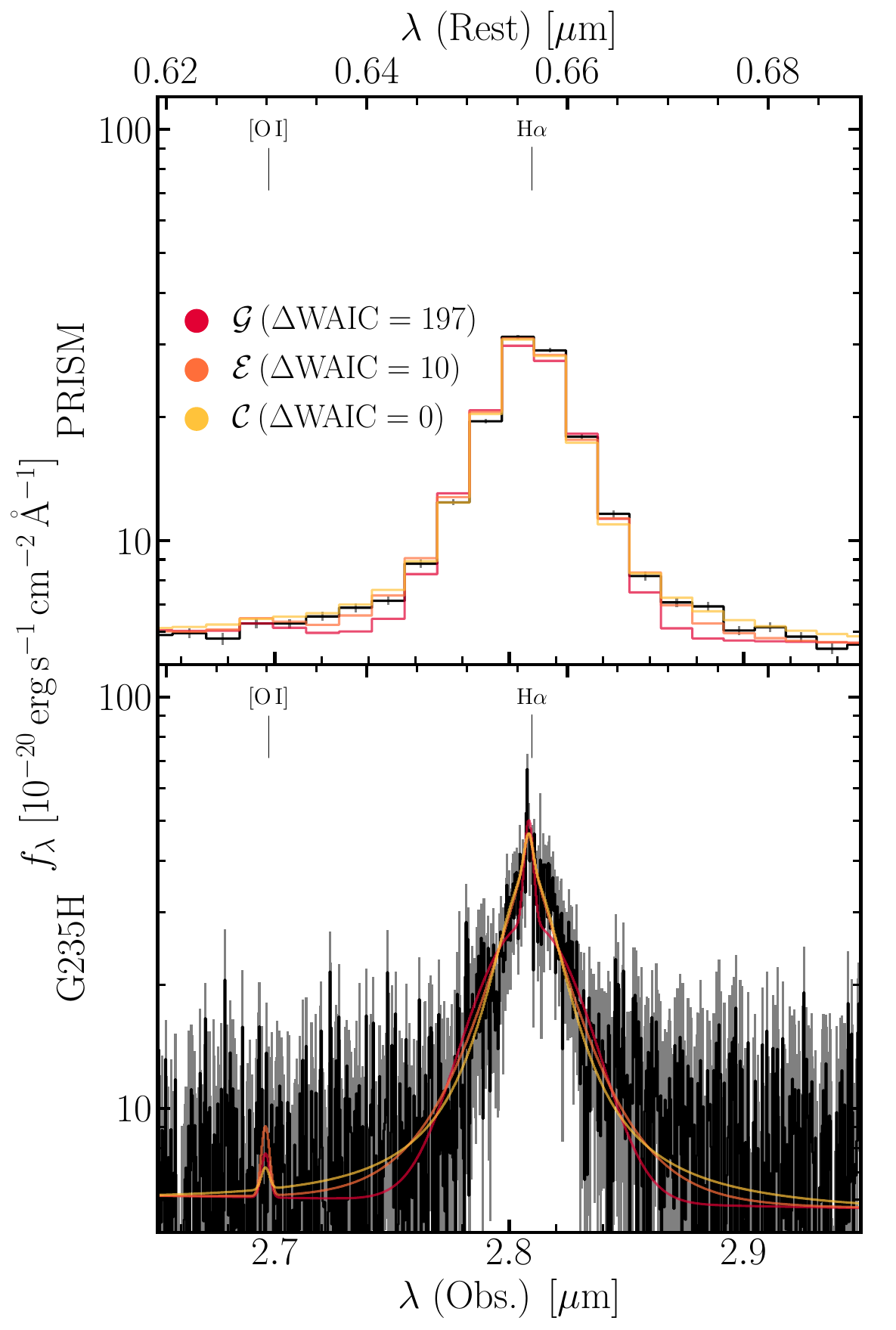}
    \caption{
    Zoom-in on the H$\alpha$ emission line in the PRISM (top) and G235H (bottom) spectra. 
    The observed data are shown in black with 1$\sigma$ uncertainties in grey. 
    We overplot the best-fit \texttt{unite} models using a Gaussian narrow plus different broad line profiles: Gaussian (red), exponential (orange), and Lorentzian (yellow). 
    The fit strongly prefers exponential and Lorentzian profiles, suggesting extended wings.
    }
    \label{fig:broad}
\end{figure}

In short, we find clear evidence for broad ($\rm FWHM \sim 2500-3200\,km\,s^{-1}$, depending on the assumed line profile) Balmer and Paschen lines (Figure \ref{fig:broad}). 
Notably, the broad emission lines are significantly better fit by Lorentzian or exponential profiles compared to a Gaussian profile (see Figure~\ref{fig:scatter} left and Appendix~\ref{app:lines}), consistent with the non-Gaussian line profiles observed in LRDs with high-quality spectroscopy \citep[e.g.,][]{labbe_UnambiguousAGNBalmer_2024,rusakov_JWSTsLittleRed_2025,degraaff_RemarkableRubyAbsorption_2025,torralba_WarmOuterLayer_2025}.
Appendix~\ref{app:lines} further details the model used, and also provides the recovered fluxes, rest equivalent widths (EWs), and kinematics. 

\begin{figure*}[ht!]
    \centering
    \includegraphics[width=0.83\textwidth]{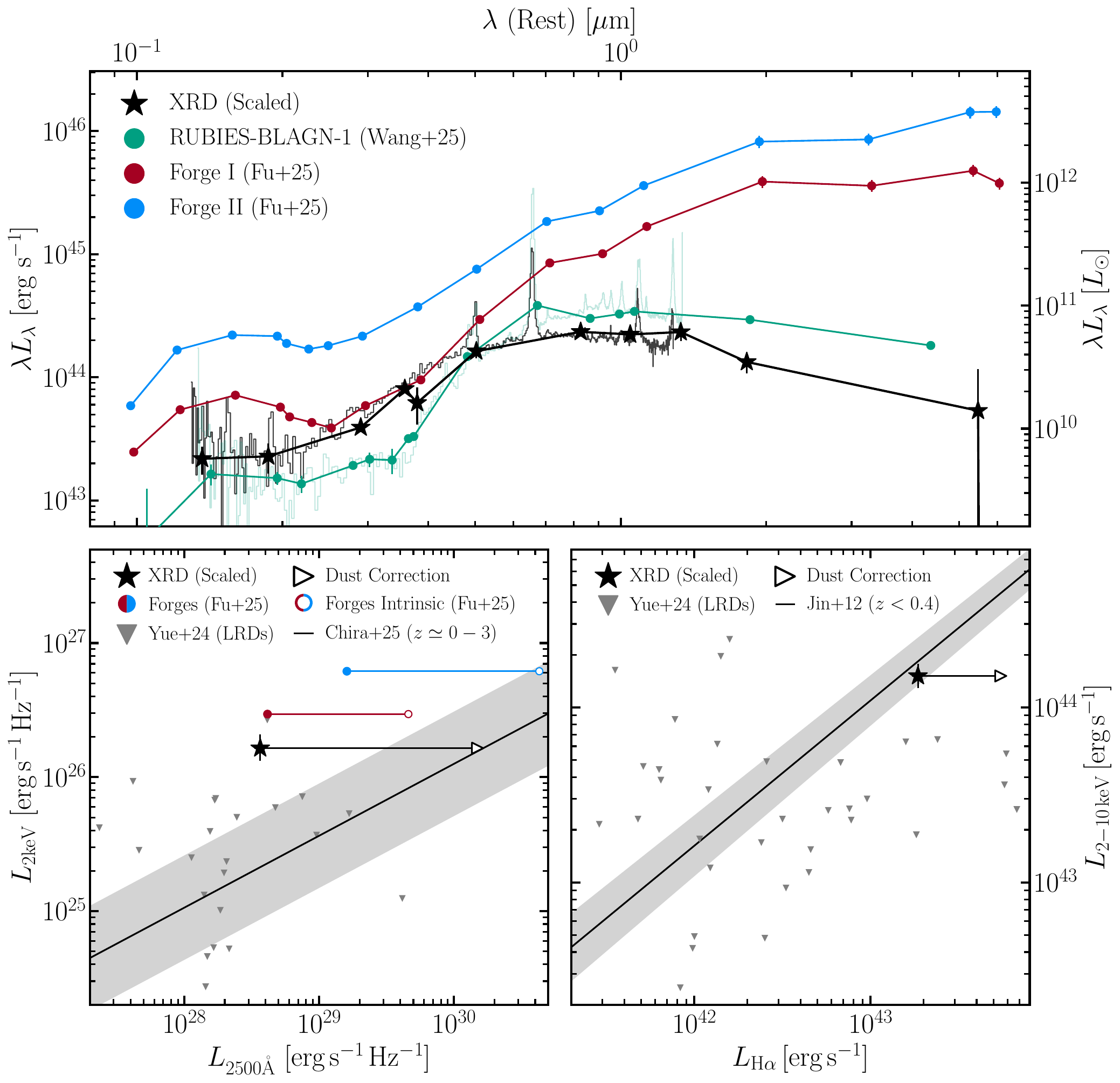}
    \caption{
        Top: The rest-frame SED of the XRD (black) compared to RUBIES-BLAGN-1 \citep[green;][]{wang_RUBIESJWSTNIRSpec_2025} and Forge I and II \citep[red and blue respectively;][]{fu_DiscoveryTwoLittle_2025}.
        Bottom Left:
        $L_{\rm 2,keV}$ versus $L_{2500\rm\mathring{A}}$ for the XRD, the Forges, and the X-ray upper limits for LRDs nondetections from \citet{yue_StackingXRayObservations_2024} (grey triangles).
        The XRD and Forge sources lie well above the standard $z\simeq0-3$ AGN relation \citep{chira_RevisitingXraytoUVRelation_2026}, indicating they are X-ray luminous for their UV emission.
        Applying a dust correction (\S \ref{ssec:continuum}) to the XRD, along with the corrections for the Forges in \citet{fu_DiscoveryTwoLittle_2025}, brings all three sources closer to the relation.
        Bottom Right: Comparison of X-ray and H$\alpha$ luminosity.
        The measured XRD properties follow the relation for $z<0.4$ unobscured AGN \citep{jin_CombinedOpticalXray_2012}, but correcting the H$\alpha$ luminosity for dust in the same manner as $L_{2500\rm\mathring{A}}$ produces an H$\alpha$ luminosity that is overluminous for its X-ray emission.
    }
    \label{fig:compare}
\end{figure*}

\subsection{X-Ray Data}\label{ssec:xray}

XDEEP2 \citep{goulding_ChandraXRayPointsource_2012} provides Chandra ACIS data from which this source’s spectrum was extracted. We stack the multi-epoch data and perform a basic spectral fitting in XSPEC \citep{arnaud_XSPECFirstTen_1996} using an attenuated power law model accounting for foreground Milky Way absorption.
The X-ray spectral analysis reveals a rather typical AGN X-ray spectrum: an intrinsic rest-frame $\rm 2 - 10\,keV$ luminosity of $\sim$$10^{44}$\,erg\,s$^{-1}$ with a typical intrinsic power-law slope of $\Gamma = 1.8$ and moderate attenuation $N_\textrm{H} \sim 10^{22}$\,cm$^{-2}$.
This places the source near the conventional boundary between unobscured (Type I) and obscured (Type II) X-ray AGN \citep{hickox_ObscuredActiveGalactic_2018}.
This high luminosity, combined with moderate evidence for X-ray variability (see Appendix~\ref{app:xray}), unambiguously identifies the source as an AGN \citep{brandt_CosmicXraySurveys_2015}.
We compare the X-ray scaling relationships of the XRD to other AGN and LRD populations in Figure~\ref{fig:compare}, which we discuss in detail in Section~\ref{sec:dust}.
Results from the XSPEC X-Ray analysis are provided in Table~\ref{tab:xrd} whereas additional details of the X-ray data and analysis are provided in Appendix~\ref{app:xray}.

\section{A Dust-Reddened AGN?}\label{sec:dust}

The XRD has unambiguous signatures of SMBH accretion, exhibiting both a high X-ray luminosity ($L_{\rm 2-10\,keV} \sim 10^{44}$\,erg\,s$^{-1}$) and broad Balmer emission lines ($\rm FWHM \sim 2700-3200\,km\,s^{-1}$).
In further, albeit tentative, support of a typical AGN origin, the rest-UV spectrum reveals emission from C\,{\sc iv} and C\,{\sc iii}], detected at $2.3\sigma$ and $6.2\sigma$ significance with measured EWs of 25 and 43\AA, respectively (Table~\ref{tab:lines}).
Such high-ionization lines are typically observed in AGN but are rare in LRDs with a few notable exceptions \citep[e.g.,][]{labbe_UnambiguousAGNBalmer_2024,akins_StrongRestUVEmission_2024}. 
Although C\,{\sc iii}] emission can originate from star formation, significant C\,{\sc iv} is challenging to produce via stellar photoionization processes alone \citep{hirschmann_SyntheticNebularEmission_2019}, potentially favoring an AGN origin despite the modest significance of the detection.
We also search for Mg\,{\sc ii}, but the line falls at the minimum of the PRISM's wavelength-dependent resolution, precluding a robust detection.

However, other features of the XRD appear remarkably LRD-like and distinct from typical dust-obscured quasars (Figure~\ref{fig:xrd}).
Specifically, the XRD has blue mid-IR colors indicative of a falling continuum which differs from obscured quasars that typically show enhanced mid-IR emission from a hot dust torus.
Instead, the XRD shares the steep red rest-optical continuum characteristic of LRDs, though its spectral inflection point occurs blueward of the Balmer limit ($\sim2000$\,\AA) unlike typical LRDs \citep{setton_LittleRedDots_2024,degraaff_LittleRedDots_2025}.

Nevertheless, the X-ray luminosity stands in stark contrast to its comparatively weak rest-UV emission, indicating that the central AGN engine must be subject to significant obscuration. In this section we aim to assess whether the XRD can be explained as a dust-reddened AGN.
We first place the XRD in the context of empirical X-ray scaling relationships.
We then test whether a dust-reddened AGN model can reproduce the broadband SED and the rest-optical spectrum, without violating the X-ray observations, mid-IR constraints, or typical dust laws for galaxies and AGN.

\subsection{X-ray Scaling Relationships}\label{ssec:scaling}

In the bottom half of Figure~\ref{fig:compare}, we compare the XRD's intrinsic, that is absorption corrected, X-ray properties to standard AGN scaling relationships.
The XRD is significantly more X-ray luminous than the upper limits for LRDs \citep{yue_StackingXRayObservations_2024}, and its X-ray luminosity is therefore comparable to that of the ``Forges'' objects \citep{fu_DiscoveryTwoLittle_2025} that were recently proposed to be transitionary sources between LRDs and quasars. 
Unlike the Forges, however, the XRD lacks the mid-IR rise attributed to dust emission seen for those sources.

All three sources lie above the standard $L_X-L_{UV}$ relation \citep{chira_RevisitingXraytoUVRelation_2026}, indicating they are remarkably X-ray luminous given their observed UV emission.
Applying a dust correction based on spectral continuum fitting (\S \ref{ssec:continuum}) to the UV luminosity would bring the source onto this relation, but applying the same correction to the H$\alpha$ luminosity would move it below the $L_X-L_{\rm H\alpha}$ relation for unobscured AGN \citep{jin_CombinedOpticalXray_2012}, which it currently follows.
This tension implies that either the broad line and continuum regions are subject to different levels of obscuration (i.e.\ the broad line region is less obscured), or that these high-redshift sources deviate from local scaling relationships, which are themselves susceptible to selection effects and intrinsic scatter.

It is also instructive to compare the XRD's bolometric output to its intrinsic X-ray luminosity: $k_{\rm X} \equiv L_{\rm bol}/L_{\rm 2-10\,keV}$. 
We integrate our sub-mm through X-ray observations to estimate $L_{\rm bol} \simeq 5\times10^{44}$\,erg\,s$^{-1}$, implying $k_X \simeq 3.3$. 
Alternatively, we can infer $L_{\rm bol}$ from the H$\alpha$ luminosity using empirical scalings: adopting a canonical \citet{greene_EstimatingBlackHole_2005} calibration gives $k_X\simeq18$, whereas adopting the LRD motivated calibration from \citet{greene_WhatYouSee_2025} calibration motivated by LRDs gives $k_X\simeq2.1$. 
The latter is much closer to the direct SED-based estimate, and together these estimates suggest the XRD is X-ray bright for its bolometric output when compared to typical AGN at similar luminosities \citep[Figure 4:][]{duras_UniversalBolometricCorrections_2020}, the opposite of what is observed in some high-$z$ samples which are typically X-ray weak \citep[e.g.,][]{maiolino_JWSTMeetsChandra_2025}.

\begin{figure*}[ht!]
    \centering
    \includegraphics[width=0.83\textwidth]{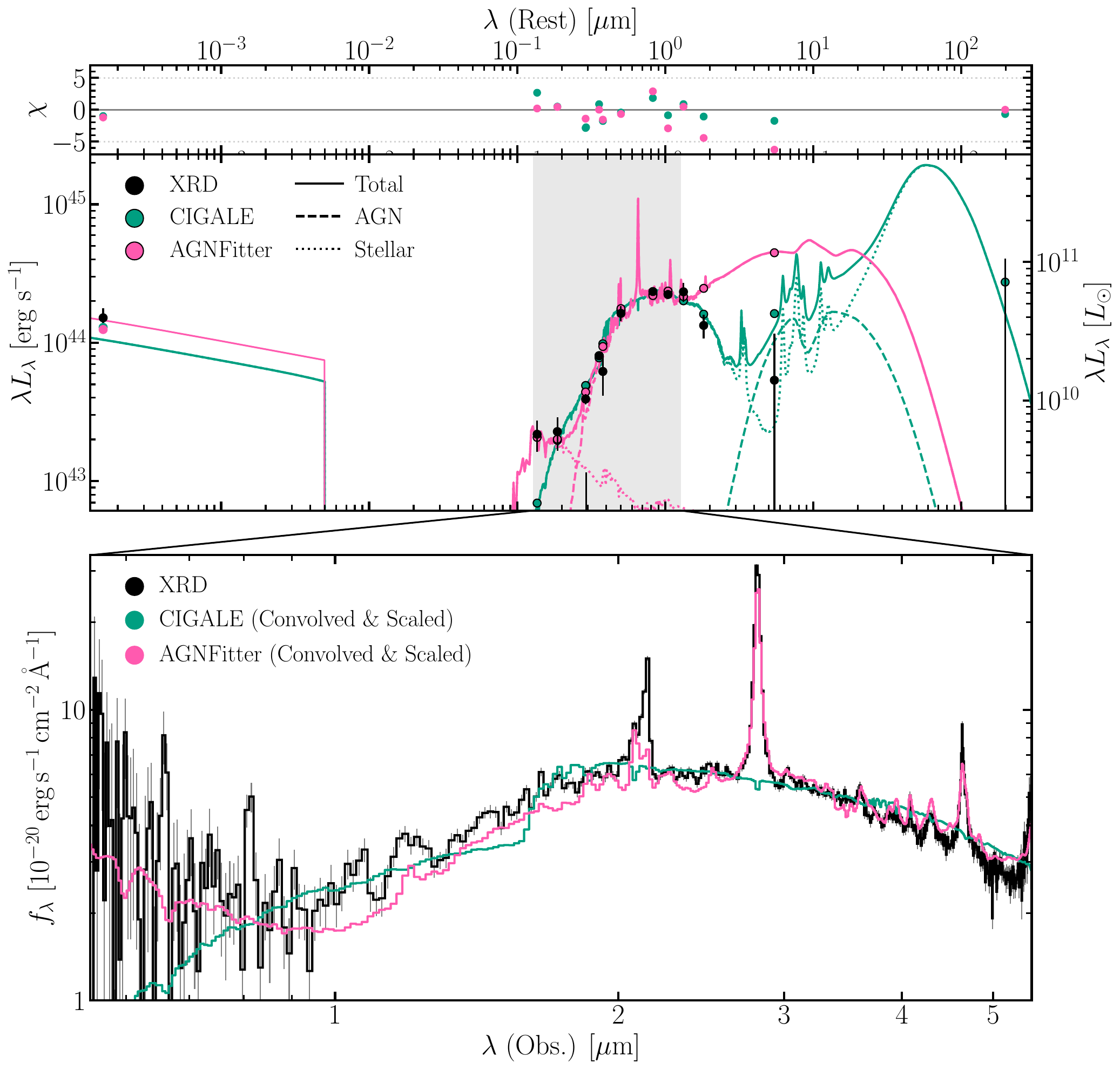}
    \caption{
    X-ray through sub-mm SED fitting of the XRD with both CIGALE (green) and AGNFitter (pink). 
    The top panel shows the best fits (solid) decomposed into their AGN (dashed) and stellar (dotted) components along with the scaled residuals. 
    The bottom panels compares the models, scaled and convolved to the PRISM resolution, to the PRISM spectrum; note that the models are not fit to the spectral data.
    In order to explain the X-ray luminosity, both models require a strong AGN component.
    However, to minimize discrepancy with the relatively weak mid-IR flux, CIGALE heavily obscures the AGN component, requiring an evolved stellar population to dominate in the rest-optical, inconsistent with the spectrum.
    AGNFitter on the other hand is able to fit the rest-optical spectrum with an AGN, but substantially overpredicts in  the mid-IR. 
    }
    \label{fig:sed}
\end{figure*}

\begin{figure*}[ht!]
    \centering
    \includegraphics[width=\textwidth]{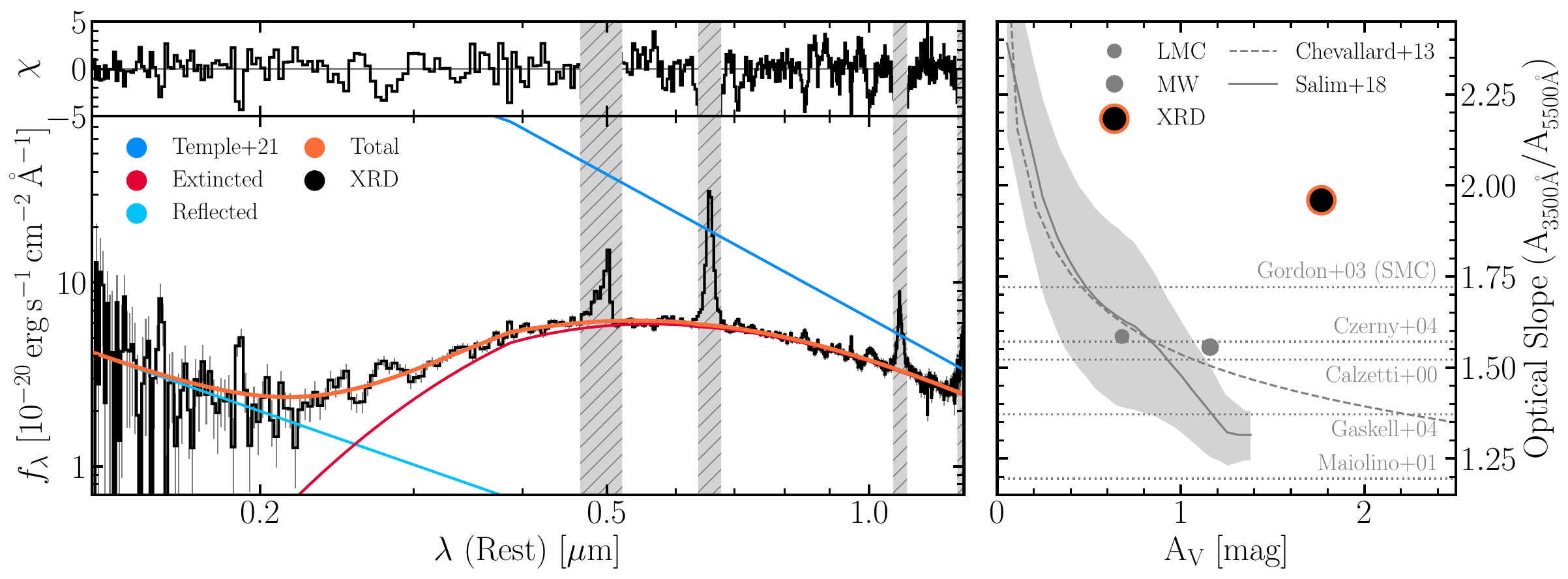}
    \caption{
    Following the approach of \citet{ma_UNCOVER404Error_2025}, we attempt to model the PRISM spectrum of the LRD assuming it is dominated by a dust-extincted, intrinsically UV-luminous, AGN. 
    We attenuate the \citet{temple_ModellingType1_2021} AGN--power-law continuum using a flexible \citet{noll_AnalysisGalaxySpectral_2009} attenuation law plus an unattenuated reflection component. 
    Although we obtain a good fit to the PRISM data, the necessary extinction ($\rm A_V \sim 1.8$) implies a Balmer decrement ($\rm H\alpha/H\beta \sim 6-7$) that is significantly smaller than the measured value ($9.5\pm0.5$).
    The left panel shows the best-fit attenuated AGN model (orange) decomposed into the intrinsic (blue), reflected (cyan), and extincted (red) components. 
    The right panel compares the derived extinction parameters to typical galaxy populations \citep{chevallard_InsightsContentSpatial_2013,salim_DustAttenuationCurves_2018}, the Large Magellanic Cloud and Milky Way \citep[LMC \& MW;][]{gordon_QuantitativeComparisonSmall_2003}, standard dust laws \citep{calzetti_DustContentOpacity_2000, gordon_QuantitativeComparisonSmall_2003}, and averages from local AGN populations \citep{maiolino_DustActiveNuclei_2001,gaskell_NuclearReddeningCurve_2004,czerny_ExtinctionDueAmorphous_2004}.
    }
    \label{fig:dust}
\end{figure*}

\subsection{SED Fitting}\label{ssec:sed}

We use two SED fitting routines to test across AGN model implementations: CIGALE \citep{burgarella_StarFormationDust_2005,noll_AnalysisGalaxySpectral_2009,boquien_CIGALEPythonCode_2019,yang_XCIGALEFittingAGN_2020} and AGNFitter \citep{calistrorivera_AGNfitterBayesianMCMC_2016,martinez-ramirez_AGNFITTERRXModelingRadiotoXray_2024}.
Both models are fit using only the photometry described in Section \ref{ssec:phot} as well as the X-ray constraints on $L_{\rm2-10\,keV}$ and $\Gamma$ detailed in Section \ref{ssec:xray}.
The full SED models are shown in Figure \ref{fig:sed} and are decomposed into their respective stellar and AGN components. 
In addition, by convolving the model with the wavelength-dependent LSF of the PRISM and multiplying the models by the measured spectroscopic scale factor, we also compare the predicted model SED to the NIRSpec spectrum, although we stress that the spectrum was not used in the fitting itself.

We adopt this photometry-only approach to keep the SED constraints focused on the broadband energy budget and the implied X-ray--to--mid-IR scaling relations, rather than on achieving an optimal fit to the densely sampled rest-optical continuum in the PRISM data.
Including the spectrum directly in the likelihood would strongly up-weight the rest-optical regime relative to the sparsely sampled mid-IR and X-ray constraints, potentially driving the fit toward models that reproduce the detailed continuum shape, although they fail to satisfy the global SED properties.
We therefore use the spectrum as an external consistency check on the SED models and perform dedicated spectral modeling in Section~\ref{ssec:continuum}.

\subsubsection{CIGALE}\label{sssec:cigale}

CIGALE is able to model the XRD photometry from the X-ray through the sub-mm and is designed to account for the energy balance for any dust-attenuated stellar components. 
We use CIGALE with the \citet{lopez_CIGALEModuleTailored_2024} X-ray prescription, which implements the $L_{\rm X}-L_{\rm12\mu m}$ scaling relationship from \citet{gandhi_ResolvingMidinfraredCores_2009}, alongside the SKIRTOR \citep{stalevski_3DRadiativeTransfer_2012,stalevski_DustCoveringFactor_2016} AGN models. 
Briefly, for the galaxy component we use a delayed-$\tau$ star-formation history, the \citet{bruzual_StellarPopulationSynthesis_2003} stellar population models, \citet{calzetti_DustContentOpacity_2000} dust attenuation law, and the THEMIS \citep{jones_GlobalDustModelling_2017} dust emission prescription. 
We provide the full details and results of our CIGALE SED fitting in Appendix \ref{app:sed} and Table \ref{tab:cigale}.

The best-fitting CIGALE model (Figure~\ref{fig:sed}) explains the red--rest-optical continuum as dominated by an older stellar component ($t_\textrm{age}\sim1\,\textrm{Gyr};\,M_\star \sim 2.5\times10^{10}\,\textrm{M}_\odot$) that is moderately dust attenuated ($\rm A_V\sim0.5$).
However, to maintain the \citet{gandhi_ResolvingMidinfraredCores_2009} scaling relationships between the X-ray and mid-IR, the model heavily suppresses the AGN contribution in the optical through mid-IR with a torus with a high covering fraction ($\theta_{\rm open}\approx55-85^\circ$; $\theta_{\rm open} = 90^\circ$ is unity covering fraction) with a direct view through the torus ($i \approx 50 - 90^\circ$; $i=0^\circ$ is face on).
The resulting torus geometry effectively hides the hot dust components at the inner torus radii, preventing the significant mid-IR emission.
Although this produces a good fit to the photometry, the resulting stellar-dominated model is strongly in tension with the PRISM spectrum due to the lack of a 4000\AA\ break as well as emission lines.

\subsubsection{AGNFitter}\label{sssec:agnfitter}

AGNFitter is primarily designed for fitting the SED of AGNs and is well suited for measuring AGN properties and determining the level of obscuration.
It achieves this by decomposing the SED into four physical components: an accretion disk, a hot dust torus, a stellar population, and cold dust in star-forming regions.
We utilize the \citet{temple_ModellingType1_2021} accretion disk models, SKIRTOR torus models, \citet{bruzual_StellarPopulationSynthesis_2003} stellar population models, and \citet{schreiber_EGGHatchingMock_2017} starburst models.
The X-ray emission is linked to the mid-IR (6\,$\mu$m) luminosity using the relation from \citet{stern_XRayMidinfraredRelation_2015}.

As shown in Figure \ref{fig:sed}, AGNFitter is able to reproduce the red optical continuum with a reddened AGN component.
In particular, the fit reproduces the PRISM spectroscopy quite accurately, including the strength of most hydrogen emission lines, without being fit directly to these data. 
However, the model predicts a significant hot dust torus component in the mid-IR to maintain the X-ray-to-mid-IR relation.
This predicted mid-IR emission significantly exceeds the observed Spitzer fluxes, with a 4.4$\sigma$ and 6.2$\sigma$ discrepancy in IRAC Ch.~4 and MIPS~24$\mu$m respectively (combined 7.6$\sigma$), indicating that the reddening mechanism in the XRD does not produce the expected thermal re-emission associated with a wide grid of torus models. 
We provide the full details and results of our AGNFitter SED fitting in Appendix \ref{app:sed}.

\subsection{Spectral Continuum Fitting}\label{ssec:continuum}

The SED fitting analysis indicates that, if the XRD is a dust-reddened AGN, its dust properties fall outside of the parameter range considered in standard SED fitting codes. 
We therefore also follow the empirical approach of \citet{ma_UNCOVER404Error_2025}, to determine which extinction law would be needed in order to explain the XRD as a reddened, intrinsically UV-luminous AGN. 
We adopt the double power-law AGN continuum model from \citet{temple_ModellingType1_2021} as the intrinsic spectrum.
This intrinsic continuum is then attenuated using the flexible attenuation law of \citet{noll_AnalysisGalaxySpectral_2009} that allows for a variable slope and extinction.
We also include an unattenuated reflection component to account for the UV-upturn in the spectrum.
We fit this model to the PRISM spectrum, masking out the regions around strong emission lines (H$\beta$, [O\,{\sc iii}], H$\alpha$, and Pa$\beta$).

The best-fit model, shown in Figure~\ref{fig:dust}, provides a good description of the rest UV-optical continuum spectrum ($\chi^2_\nu = 2.53$).
However, this fit requires a high extinction of $\rm A_V \sim 1.8$\,mag and an extremely steep extinction curve.
Following \citet{degraaff_RemarkableRubyAbsorption_2025}, we quantify the steepness of the attenuation curve using the optical slope, defined as the ratio of the attenuation at 3500\,\AA\ and 5500\,\AA\ ($\rm A_{3500}/A_{5500}$).
As shown in the right panel of Figure~\ref{fig:dust}, the required optical slope for the XRD is substantially steeper than standard laws for the Milky Way, Large Magellanic Cloud, Small Magellanic Cloud, and typical galaxy populations \citep{calzetti_DustContentOpacity_2000, gordon_QuantitativeComparisonSmall_2003}.
It is also significantly steeper than the average extinction curves derived for AGN \citep[e.g.,][]{maiolino_DustActiveNuclei_2001, gaskell_NuclearReddeningCurve_2004, czerny_ExtinctionDueAmorphous_2004}, which are typically shallower than the SMC law.

Such a combination of a steep optical slope and high attenuation is physically difficult to explain with dust; that is, systems with more extinction typically exhibit greyer attenuation curves.
The location of the XRD in this parameter space of high $\rm A_V$ and steep slope is therefore highly unusual and potentially nonphysical for a dust-screen scenario.

The Balmer decrement ($\rm H\alpha/H\beta$) provides an independent constraint on whether the red continuum can be explained by dust extinction.
For our statistically preferred broad-line profile models (Lorentzian and exponential), the narrow-line Balmer decrement is consistent with Case~B recombination within the uncertainties (Table~\ref{tab:lines}), suggesting little reddening in the narrow-line region and/or host galaxy.
In contrast, the measured broad-line decrement ($9.5\pm0.5$) is strongly elevated, potentially indicative of heavy nuclear dust obscuration.
However, estimating a dust extinction from this value is not straightforward as intrinsic broad-line Balmer decrements are not expected to follow Case~B: local unobscured AGNs often show enhanced intrinsic broad-line decrements up to 4.2 \citep{dong_BroadlineBalmerDecrements_2008} with models predicting values up to $\sim 5$ \citep{korista_WhatOpticalRecombination_2004}, plausibly due to high gas densities and optical-depth effects.
Thus, while the intrinsic baseline is uncertain, the elevated broad-line decrement remains formally consistent with a nuclear dust obscuration scenario.

Nevertheless, these discrepancies, namely the extreme dust properties required to explain the reddened rest-optical yet weak mid-IR emission, mirror the arguments presented in \citet{ma_UNCOVER404Error_2025} and \citet{degraaff_RemarkableRubyAbsorption_2025} against a pure dust interpretation for LRDs.

\begin{figure*}[ht!]
    \centering
    \includegraphics[width=\textwidth]{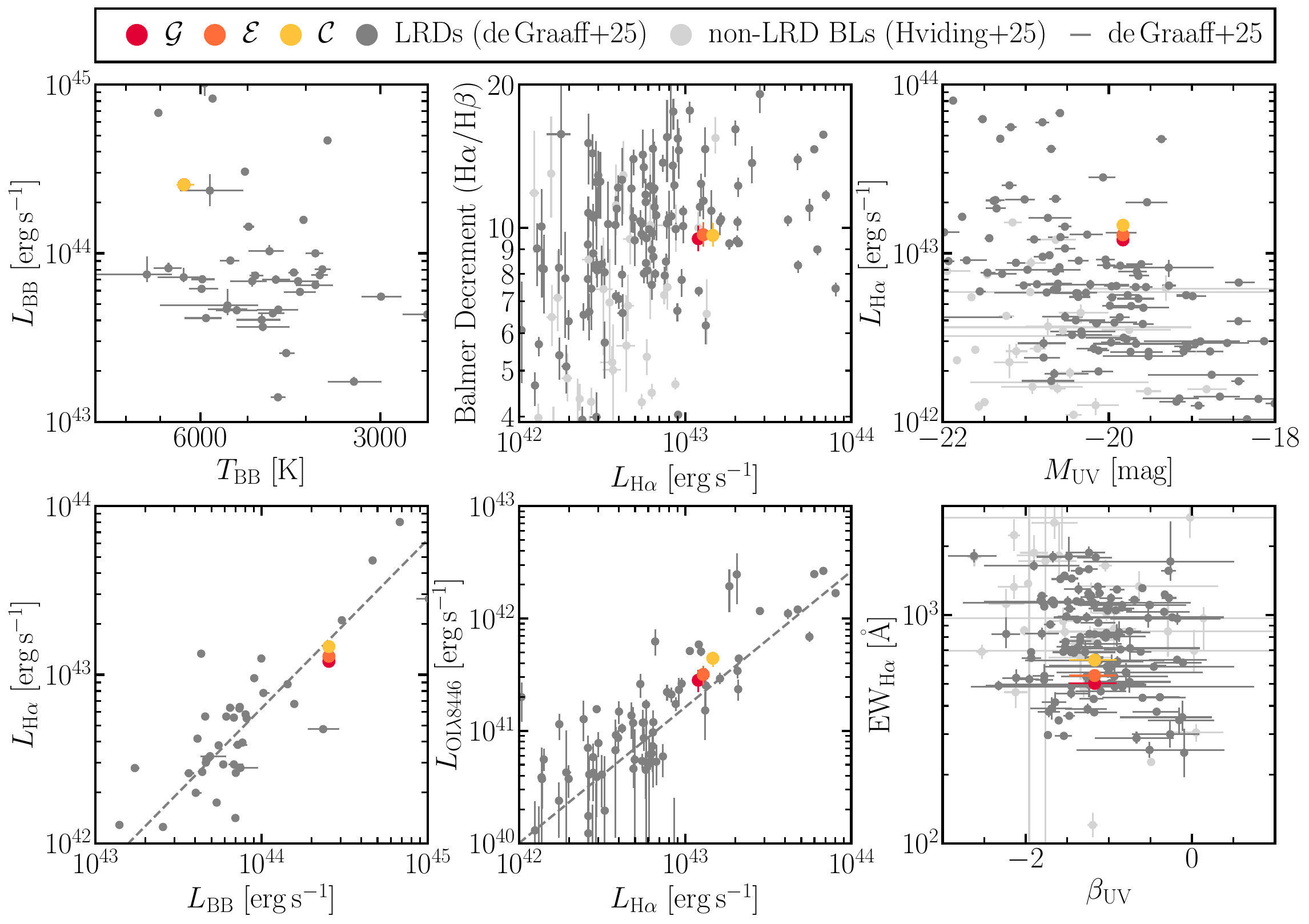}
    \caption{
    Comparison of the XRD to the population of LRDs (dark gray) and non-LRD broad-line objects (light gray) from \citet{degraaff_LittleRedDots_2025} based on emission line and modified black-body fitting.
    Panels show: Blackbody temperature vs.\ luminosity (top left); Balmer decrement vs.\ H$\alpha$ luminosity (top middle); H$\alpha$ luminosity vs.\ UV magnitude (top right); H$\alpha$ luminosity vs.\ Blackbody luminosity (bottom left); O\,{\sc i}\,$\lambda$8446\AA\ luminosity vs.\ H$\alpha$ luminosity (bottom middle); H$\alpha$ EW vs.\ UV power-law slope (bottom right).
    The XRD generally lies in regions occupied by LRDs and not non-LRD broad line galaxies and on the relations derived in \citet{degraaff_LittleRedDots_2025}.
    }
    \label{fig:scatter}
\end{figure*}

\section{Gas-Dominated Reddening?}\label{sec:gas}

The high X-ray intrinsic luminosity of the XRD ($L_{\rm 2-10\,keV} \sim 10^{44}$\,erg\,s$^{-1}$) at face value implies a correspondingly high intrinsic UV luminosity. 
As demonstrated in Section~\ref{sec:dust}, typical dust attenuation scenarios fail to reconcile this intrinsic emission with the observed red rest-optical continuum without invoking extreme dust properties. 
We therefore explore the alternative scenario proposed for the LRD population: that the intrinsic UV-continuum is reddened by dense, optically thick, gas \citep[e.g.,][]{inayoshi_ExtremelyDenseGas_2025,kido_BlackHoleEnvelopes_2025,ji_BlackTHUNDERNonstellarBalmer_2025,naidu_BlackHoleStar_2025,degraaff_RemarkableRubyAbsorption_2025}.

\subsection{The XRD in Context of the LRD Population}

Following \citet{degraaff_LittleRedDots_2025}, we decompose the PRISM spectrum into a modified blackbody ($f_\nu \propto \nu^{\beta} B_\nu$ or equivalently $f_\lambda \propto \lambda^{-\beta} B_\lambda$) representing the thermalized gas emission, and a power-law to reproduce the UV ($f_\lambda \propto \lambda^\beta$).
This simple model provides an excellent description of the XRD's spectral shape ($T_{\rm eff} \simeq 6400\,\textrm{K};\,\beta\simeq-1$), capturing both the steep red continuum and the UV upturn.
However, the derived $\beta_{BB} = -1$ indicates a continuum shape that is broader than a single-temperature blackbody, which, although consistent with typical LRDs, argues against the emission arising from a single purely thermal component.

\begin{figure*}[ht!]
    \centering
    \includegraphics[width=0.83\textwidth]{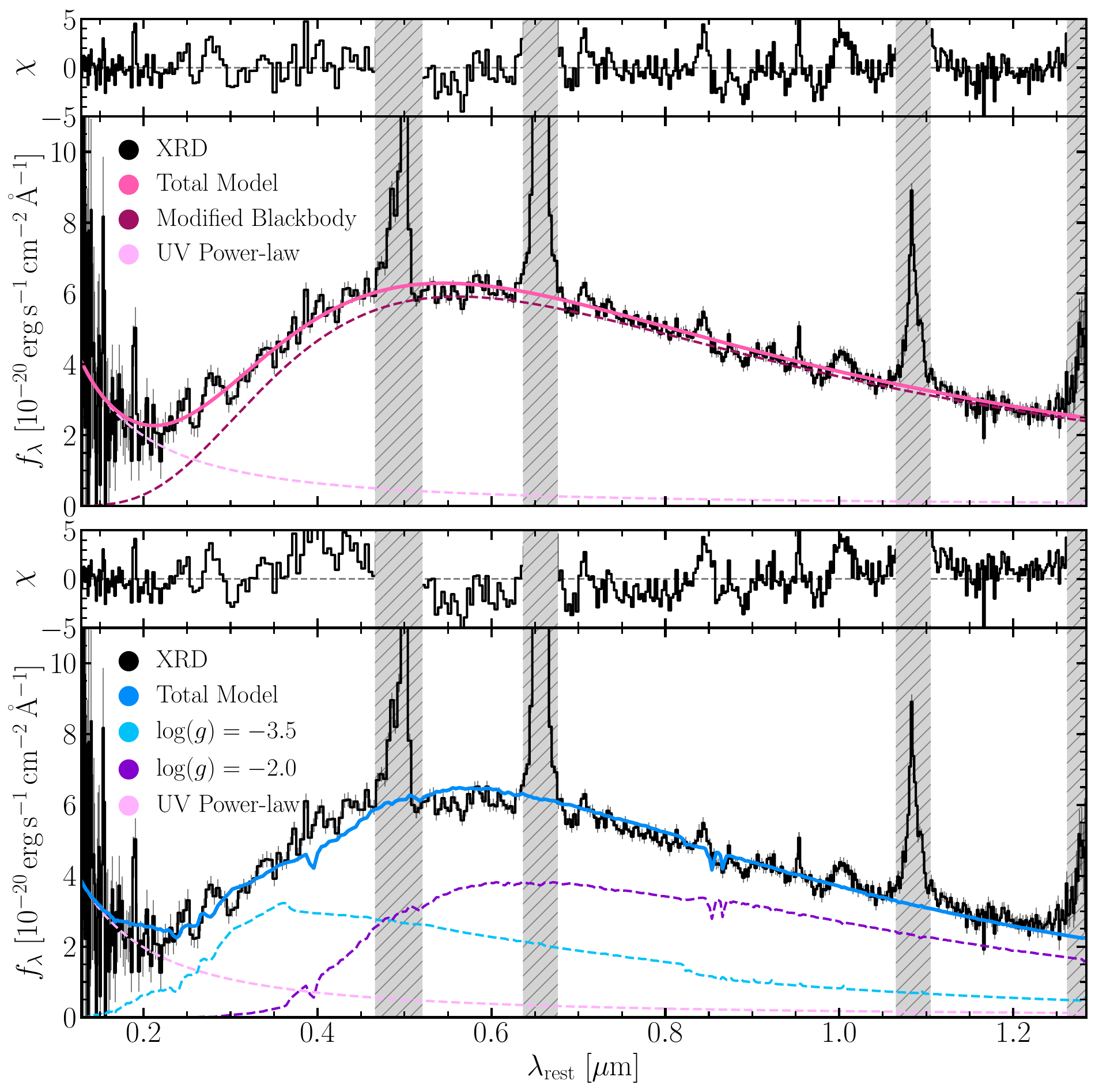}
    \caption{
    Modeling the XRD continuum as emission from a dense gas envelope.
    \textbf{Top:} Following \citet{degraaff_LittleRedDots_2025} we perform a phenomenological fit decomposing the spectrum into a modified blackbody (magenta dashed), representing the thermalized gas emission, and a UV power-law (pink dashed).
    This simple model provides an excellent description of the continuum of the XRD.
    \textbf{Bottom:} A physical fit (blue) combining two $T_{\rm eff}=5000$\,K low-density gas atmosphere models with different densities: $\log(g) = -3.5$ (cyan dashed) and $\log(g) = -2.0$ (purple dashed) that are attenuated by SMC dust ($\rm A_V\sim0.6$) plus a UV power-law (pink dashed.).
    The residuals for both fits are shown in the upper sub-panels.
    }
    \label{fig:gas}
\end{figure*}

We compare the XRD's modified blackbody parameters, and emission-line EWs and luminosities with the sample of LRDs and broad-line AGN from \citet{degraaff_LittleRedDots_2025} and \citet{hviding_RUBIESSpectroscopicCensus_2025} in Figure~\ref{fig:scatter} (for reference, the modified blackbody decomposition is shown in Figure~\ref{fig:gas}).
Regardless of the assumed broad-line profile (Gaussian, Lorentzian, or exponential), the XRD consistently falls within the region traced by the LRD population \citep{degraaff_LittleRedDots_2025}.

Specifically, the XRD exhibits an extreme Balmer decrement ($\rm H\alpha/H\beta \sim 10$; Figure~\ref{fig:scatter}, top center), comparable to that seen in the LRD population \citep{degraaff_LittleRedDots_2025}.
It also has a very large H$\alpha$ equivalent width ($\rm EW_{H\alpha} \sim 550-620$\,\AA) and is unusually H$\alpha$ luminous for its absolute UV magnitude, lying even further from the standard AGN relation than the typical LRDs (Figure~\ref{fig:scatter}, right).
Moreover, we detect strong O\,{\sc i}\,$\lambda$8446\AA\ (Table~\ref{tab:lines}) emission that follows the $L_{\rm H\alpha}-L_{\rm OI}$ scaling relation in \citet{degraaff_LittleRedDots_2025}, implying high-density, optically thick gas (Figure~\ref{fig:scatter}, bottom middle). Finally, the XRD tracks LRD-specific trends such as the correlation between H$\alpha$ luminosity and blackbody luminosity, further supporting its association with the gas-enshrouded LRD class (Figure~\ref{fig:scatter}, bottom left).

\subsection{A Late-Stage LRD?}

The fact that the UV-optical spectral properties of the XRD closely resemble those of the LRD population raises major questions regarding the physical conditions of the gas surrounding the central engine. 
Recent efforts to model LRDs \citep[e.g.,][]{liu_BalmerBreakOptical_2025,ji_BlackTHUNDERNonstellarBalmer_2025,naidu_BlackHoleStar_2025,degraaff_RemarkableRubyAbsorption_2025,taylor_CAPERSLRDz9GasEnshrouded_2025} all invoke a dense gas envelope that is Compton thick, forming a key argument in explaining their observed X-ray weakness. 
Moreover, these models give rise to a dominant thermal component in the rest-frame optical, well approximated by a $\sim$5000\,K blackbody \citep{degraaff_LittleRedDots_2025,umeda_BlackHoleEnvelopeInterpretation_2025}, while simultaneously producing strong Balmer breaks and high intrinsic Balmer decrements.

The XRD differs from these models in three key respects: it is X-ray luminous, its UV-optical SED is broader than a single thermal component, and its continuum turnover occurs blueward of the Balmer limit.
Motivated by this, we explore whether a dense envelope can still dominate the observed UV-optical emission while permitting X-rays to escape.
We adopt a patchy, multi-phase gas envelope with a sub-unity covering fraction, which can be physically interpreted as a dynamically unstable, transitional configuration.
In this picture, optically thick regions thermalize and shape the UV-optical spectrum, whereas optically thin channels provide direct sightlines to the compact X-ray source.
Although idealized, such a short-lived phase could plausibly be observed given the abundance of LRDs \citep[$\sim0.5$\,arcmin$^{-2}$;][]{kokorev_CensusPhotometricallySelected_2024} and their small inferred lifetimes \citep[$\sim10$\,Myr;][]{santarelli_EvolutionaryTracksSpectral_2025,begelman_LittleRedDots_2026}.

To explore the properties of such an envelope we use the theoretical spectral templates of H. Liu et al.\ (in prep.), which update the models in \citet{liu_BalmerBreakOptical_2025}.
Emission features of the model spectra are mainly determined by the effective temperature and the atmosphere density, the latter parameterized by $\log(g)$. 
The spectral similarities between the XRD and LRDs motivate us to start from a single template of $T_{\rm eff}=5000{\rm~K}$ and $\log(g)=-2.0$ that well describes the optical--to--near-IR SED of LRDs such as UNCOVER-45924 \citep{labbe_UnambiguousAGNBalmer_2024}. 
However, this model fails to fit the XRD continuum, producing an overly narrow and red SED. 
This calls for an additional, bluer emission component, as already suggested by the high $T_{\rm eff}$ and $\rm \beta_{BB}\sim -1$ in our modified blackbody fit.

We then test whether a hotter component could provide the required blue emission by adding templates with higher $T_{\rm eff}$ at comparable or higher $\log(g)$.
However, these models introduce a pronounced Balmer break that is not present in the NIRSpec/PRISM spectrum. 
We instead vary the atmosphere density: lower-$\log(g)$ templates naturally produce bluer continua without invoking higher temperatures.
In Figure~\ref{fig:gas} (bottom), a combination of two $T_{\rm eff}=5000\,{\rm K}$ components at different densities ($\log(g)=-3.5$ and $\log(g)=-2.0$) reproduces the broad, blue rest-optical SED while maintaining a weak Balmer break, after attenuation by moderate SMC-like dust ($A_V\sim0.6$). 
We invoke an unattenuated power law to produce the rest-UV that could arise from the host galaxy, an AGN accretion disk, or a combination of both.

We interpret this fit as evidence for a patchy, multi-phase gas envelope surrounding the black hole. 
In this picture, the blue emission and suppressed Balmer break in the low-density component ($\log(g)=-3.5$) are a consequence of scattering-dominated radiative transfer. 
We caution, however, that the solution requires fine-tuning the density to avoid an inverse Balmer break and still necessitates some dust attenuation. 
Moreover, the H. Liu et al.\ (in prep.) framework assumes an optically thick medium ($\tau \gg 1$) and may therefore not capture the true physical state of the XRD if it is transitioning between optically thick and optically thin. 
Nevertheless, this exercise highlights that the dense, optically thick atmosphere conditions inferred for typical LRDs (where the model assumptions are more likely to be valid) struggle to reproduce the XRD SED, pointing to lower densities as a potential path forward.

Finally, we note that \citet{fu_DiscoveryTwoLittle_2025} recently proposed a similar qualitative picture for the Forges, two X-ray luminous, compact red sources.
However, the XRD differs from these objects in important ways: we do not find evidence for hot dust out to rest-frame $5\,\micron$ (Figure~\ref{fig:compare}), the optical luminosity is comparable to that of typical LRDs, and the host galaxy is faint. 
Therefore, if truly reddened by gas rather than dust, the XRD offers a crucially different scenario for the late-stage phase of LRDs.

\section{Summary and Discussion}\label{sec:discussion}

The XRD presents a unique set of properties that challenge standard classification. 
Although its high X-ray luminosity ($L_{\rm 2-10\,keV} \gtrsim 10^{44}$\,erg\,s$^{-1}$) and spectral shape ($\Gamma \sim 1.8$) would normally unambiguously identify it as a typical AGN, it shares striking similarities with the LRD population.
Specifically, it exhibits a V-shaped spectral continuum with blackbody-like emission, non-Gaussian broad lines with significant wings, and weak mid-IR emission in conflict with the existence of a hot-dust torus.
However, a key distinction in the UV-optical SED remains: the spectral inflection point lies blueward of the Balmer limit, rather than at or redward of the Balmer limit as is typical of LRDs \citep{degraaff_LittleRedDots_2025}.

We have systematically explored whether the properties of the XRD can be explained by dust obscuration.
However, standard dust models fail to reproduce the observed SED without invoking a combination of an extremely steep attenuation curves at high dust extinction that are inconsistent with those observed in local galaxies or AGN. 
Dust-obscured AGN models also require hot dust components that are incompatible with the observed Spitzer photometry based on typical AGN scaling relations. 
Although hot-dust--deficient quasars have been observed in X-ray AGN surveys \citep{hao_HotdustpoorType1_2010}, they are typically UV-luminous.
The fact that the narrow-line Balmer decrement is consistent with case B recombination implies that the obscuration is primarily nuclear.
This would necessitate a nuclear dust component, which typically produces strong hot dust emission, and is therefore in tension with the absence of IR emission in a hot-dust--deficient quasar interpretation.
Although we cannot strictly rule out exotic dust scenarios for this single source without further data, for these reasons we do not prefer attributing the XRD's properties to a dust-obscured AGN.

Alternatively, we attempt to describe the continuum as arising from a dense gas envelope, consistent with the black hole star scenario proposed for LRDs.
We find that such a gas-reddened model can approximately reproduce the correct spectral shape, but it requires fine-tuning the densities of at least two optically thick components.
This suggests that the XRD may reside in a physical regime not well-captured by current static atmosphere models that assume high optical depths ($\tau \gg 1$).
The XRD may therefore represent an object in transition, where the gas envelope is multi-phase and the optical depth is intermediate between the optically thick limit of atmospheric models and the optically thin limit of photoionization codes like \texttt{cloudy} \citep{gunasekera_2025ReleaseCloudy_2025}.
Nevertheless, it remains a challenge to envision a configuration in which the optical emission is still dominated by an optically thick envelope while allowing optically thin sightlines to the compact X-ray source, yet simultaneously preventing UV emission from leaking out from the accretion disk.
This tension could be resolved by moving from spherical symmetry to non-spherical or clumpy geometries that allow direct views of the X-ray corona while maintaining high reprocessing of the accretion disk emission.

If the XRD is indeed an object in transition, it offers a unique laboratory to study the nature of LRDs.
Future observations will be critical to test this hypothesis.
JWST/MIRI spectroscopy and imaging are therefore crucial to characterize the mid-IR dust emission and determine if a hidden hot dust component exists, whereas upcoming NOEMA observations (PI: Hviding) will constrain the cold dust budget.
Although tentative X-ray variability supports the transitionary hypothesis, continued X-ray monitoring is required to confirm this, whereas NIRCam imaging would provide constraints on any long-term rest-optical changes.

Crucially, if the rest-optical continua of LRDs are indeed dominated by dense, optically thick, gas envelopes and the Balmer emission lines are broadened primarily by scattering processes, as opposed to gravitational motions, then there remain no direct pieces of observational evidence that the luminosity of LRDs are powered by accretion onto a SMBH. 
Confirming the XRD as a transitionary LRD would provide smoking-gun evidence for SMBH accretion at the center of at least some fraction, if not all, of this population. 
Furthermore, the optically thin sightlines down into the central engine of the XRD should exhibit more short-timescale variability than the rest of the LRD population and therefore offer critical constraints on the accretion physics and even black hole masses of the entire LRD population. 

\begin{acknowledgments}

We thank Bernd Husemann for his critical contributions to the NIRSpec Wide GTO Survey, and in particular his help in selecting high-priority X-ray luminous targets. \

REH acknowledges support by the German Aerospace Center (DLR) and the Federal Ministry for Economic Affairs and Energy (BMWi) through program 50OR2403 `RUBIES'.
AdG acknowledges support from a Clay Fellowship awarded by the Smithsonian Astrophysical Observatory.
AJB acknowledges funding from the ``FirstGalaxies'' Advanced Grant from the European Research Council (ERC) under the European Union's Horizon 2020 research and innovation program (Grant agreement No. 789056).
RPN thanks Neil Pappalardo and Jane Pappalardo for their generous support of the MIT Pappalardo Fellowships in Physics.
Support for this work was provided by The Brinson Foundation through a Brinson Prize Fellowship grant.
H\"U acknowledges funding by the European Union (ERC APEX, 101164796). 
Views and opinions expressed are however those of the authors only and do not necessarily reflect those of the European Union or the European Research Council Executive Agency. 
Neither the European Union nor the granting authority can be held responsible for them.
GV acknowledges support from European Union’s HE ERC Starting Grant No. 101040227 - WINGS.
BW acknowledges support provided by NASA through Hubble Fellowship grant HST-HF2-51592.001 awarded by the Space Telescope Science Institute, which is operated by the Association of Universities for Research in Astronomy, In., for NASA, under the contract NAS 5-26555.

The data products presented herein were retrieved from the Dawn JWST Archive (DJA). 
DJA is an initiative of the Cosmic Dawn Center (DAWN).

This work is based in part on observations made with the NASA/ESA/CSA James Webb Space Telescope. 
The data were obtained from the Mikulski Archive for Space Telescopes at the Space Telescope Science Institute, which is operated by the Association of Universities for Research in Astronomy, Inc., under NASA contract NAS 5-03127 for JWST. 
These observations are associated with programs numbers GTO-1213.
The data described here may be obtained from the MAST archive at
\dataset[doi:10.17909/qffz-b324]{https://dx.doi.org/10.17909/qffz-b324}.

This work is based on observations taken by the 3D-HST Treasury Program (GO 12177 and 12328) with the NASA/ESA HST, which is operated by the Association of Universities for Research in Astronomy, Inc., under NASA contract NAS5-26555.

This work makes use of color palettes created by Martin Krzywinski designed for colorblindness. 
The color palettes and more information can be found at \url{http://mkweb.bcgsc.ca/colorblind/}.
\end{acknowledgments}

\facilities{CXO (ACIS), HST (ACS, WFC3), CFHT (WIRCam), JWST (NIRSpec), Spitzer (IRAC, MIPS), JCMT (SCUBA)}

\software{
\texttt{Astropy} \citep{astropycollaboration_AstropyCommunityPython_2013,astropycollaboration_AstropyProjectBuilding_2018,astropycollaboration_AstropyProjectSustaining_2022},
\texttt{dust\_attenuation},
\texttt{dust\_extinction} \citep{gordon_Dust_extinctionInterstellarDust_2024},
\texttt{jax} \citep{bradbury_JAXComposableTransformations_2018},
\LaTeX\ \citep{lamport_LaTeXDocumentPreparation_1994}, 
\texttt{Matplotlib} \citep{hunter_Matplotlib2DGraphics_2007},
\texttt{NumPy} \citep{oliphant_GuideNumPy_2006,vanderwalt_NumPyArrayStructure_2011, harris_ArrayProgrammingNumPy_2020},
\texttt{NumPyro} \citep{phan_ComposableEffectsFlexible_2019}, 
\texttt{scipy} \citep{virtanen_SciPy10Fundamental_2020},
\texttt{sedpy} \citep{johnson_BdjProspectorInitial_2017},
\texttt{specutils} \citep{astropy-specutilsdevelopmentteam_SpecutilsSpectroscopicAnalysis_2019},
\texttt{unite} \citep{hviding_TheSkyentistUniteVersion_2025}.
}

\appendix
\counterwithin{figure}{section}
\counterwithin{table}{section}

\section{X-Ray Modeling} \label{app:xray}

We utilize the Chandra ACIS data from the DEEP2 survey \citep{goulding_ChandraXRayPointsource_2012}. 
The source was observed between dates 2005-03-24 and 2008-06-23 for a total of 29 individual observations. 
Photons were extracted following the procedure outlined in \citet{goulding_ChandraXRayPointsource_2012} of the astrometry-corrected merged Level 2 events file and from a circular aperture with radius consistent with the 90\% encircled energy fraction of the combined point spread function. 
The background region was contemplated from an annulus with inner radius of $1.3\times$ the PSF radius and outer radius with radius of $5\times$ the PSF radius excluding the presence of any nearby bright sources. 
Necessary instrument response files were extracted and combined from the individual obsids.

The spectrum was extracted from the stacked multi-epoch observations to maximize the signal-to-noise ratio. 
We performed spectral fitting using XSPEC \citep{arnaud_XSPECFirstTen_1996} minimizing the Cash statistic \citep{cash_ParameterEstimationAstronomy_1979} with a model consisting of an absorbed power-law with both Galactic and intrinsic absorption components: \texttt{model = Nh\_MW * (Nh * powerlaw)\_src}. 
The Galactic absorption column density was fixed to the known $N_{\rm H, MW} = 9.25 \times 10^{19}$\,cm$^{-2}$. 
The fit results, summarized in Table~\ref{tab:xrd}, indicate a typical AGN spectrum with a photon index of $\Gamma \approx 1.8$ and an intrinsic luminosity of $L_{\rm 2-10\,keV} \approx 1.5 \times 10^{44}$\,erg\,s$^{-1}$. 
The source shows moderate intrinsic absorption with $N_{\rm H} \approx 2.4 \times 10^{22}$\,cm$^{-2}$.
Although the individual measurements have significant uncertainties, together they indicate weak evidence for variability, specifically the null hypothesis is rejected at the 90\% ($1.6\sigma$) level, potentially similar to the X-ray variability seen in \citet{fu_DiscoveryTwoLittle_2025}.
Although the individual measurements have significant uncertainties, together they indicate weak evidence for variability, specifically the null hypothesis is rejected at the 90\% ($1.6\sigma$) level, potentially similar to the X-ray variability seen in \citet{fu_DiscoveryTwoLittle_2025}.

\begin{figure*}[ht!]
    \centering
    \includegraphics[width=\textwidth]{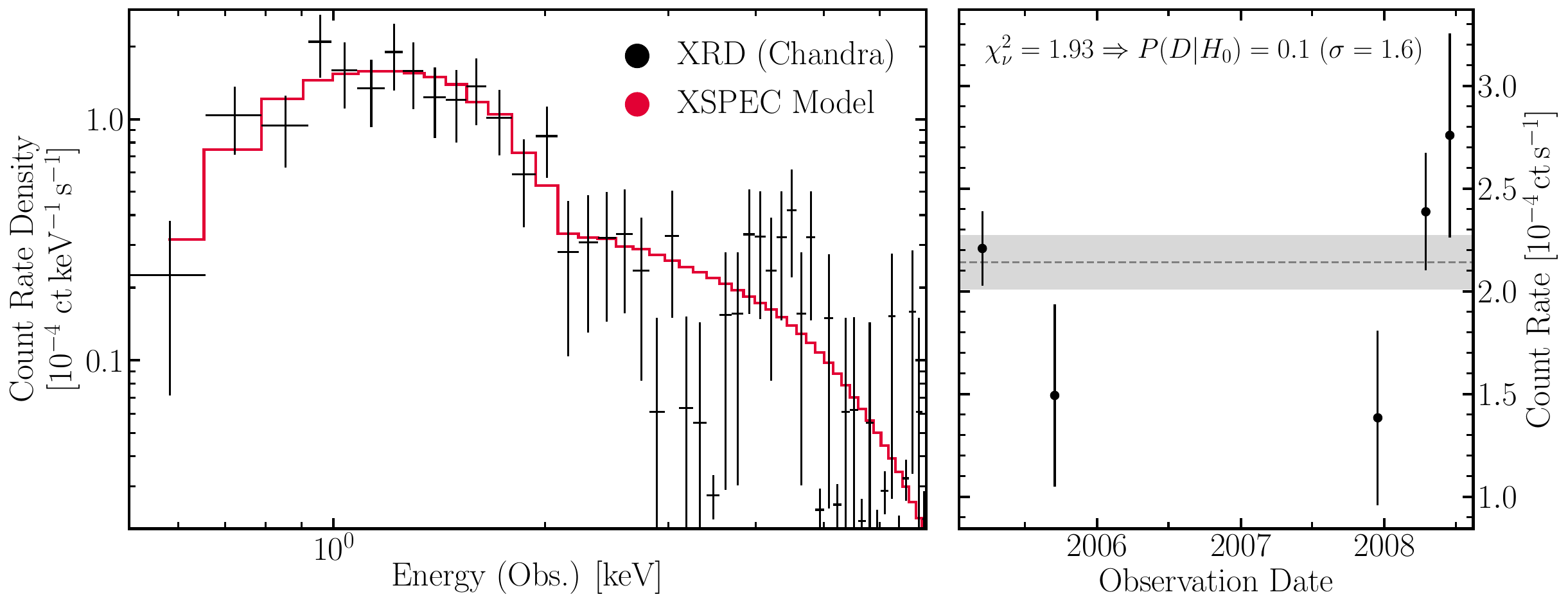}
    \caption{
    X-ray properties of the XRD. Left: The Chandra ACIS-I X-ray spectrum (black) with the best-fit absorbed power-law model (red). 
    The spectrum is consistent with a typical AGN ($L_\textrm{2$-$10\,keV}\sim10^{44}$\,erg\,s$^{-1}$ and $\Gamma\sim1.8$) with moderate absorption ($N_{\rm H} \sim 10^{22}$\,cm$^{-2}$). 
    Right: The long-term X-ray light curve in the observed 0.5$-$7\,keV band. 
    The dashed line and gray region show the weighted mean and error on the weighted mean respectively. 
    The sources shows weak evidence for variability, specifically the null hypothesis is rejected at the 90\% ($1.6\sigma$) level. 
    }
\end{figure*}

\section{Emission Line Fitting} \label{app:lines}

We perform a detailed emission line analysis using the \texttt{unite} \citep{hviding_TheSkyentistUniteVersion_2025} spectral fitting code. 
A key feature of our analysis is the simultaneous fitting of the PRISM, G235H, and G395H spectra. 
This approach allows us to leverage the high signal-to-noise of the PRISM data to constrain the continuum and broad line wings while using the high-resolution grating data to resolve narrow line components and kinematic substructure. 
The fitting procedure explicitly accounts for the wavelength-dependent line-spread function (LSF) of each disperser and integrates each pixel to account for the critical undersampling of the NIRSpec LSF.
We use the NIRSpec LSF curves of an idealized point source obtained with \texttt{msafit} \citep{degraaff_IonisedGasKinematics_2024}. 

\begin{figure*}[ht!]
    \centering
    \includegraphics[width=\textwidth]{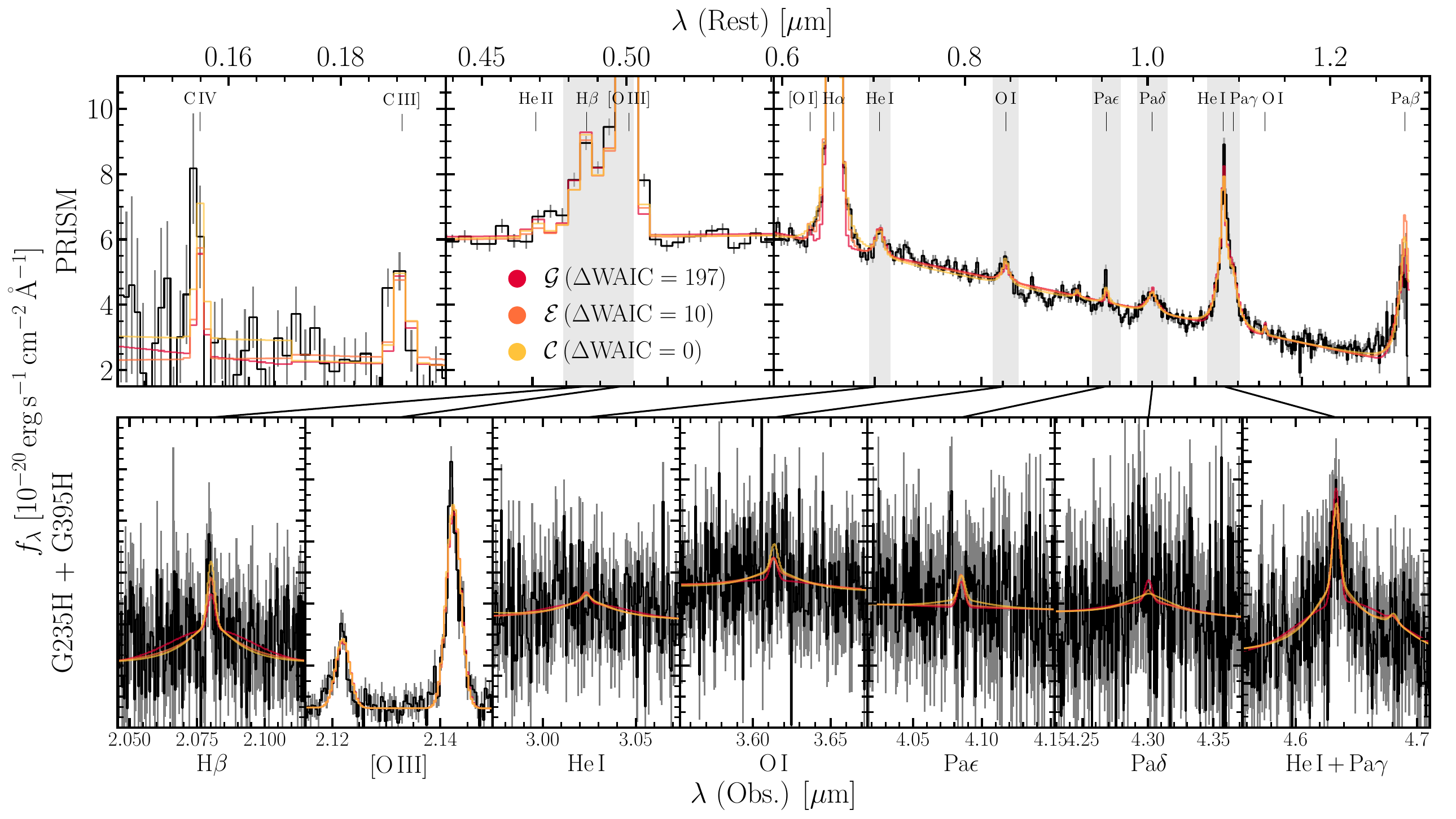}
    \caption{
    Emission line fits to the XRD with \texttt{unite} leveraging all three dispersers simultaneously while accounting for wavelength dependence and undersampling of the line-spread functions.
    All lines are fit with a narrow component with a common width. Permitted transitions are fit with an additional broad Gaussian (red), Lorentzian (orange), or exponential (yellow) component.
    The top panels show the PRISM spectrum, whereas the bottom show the grating spectra.
    The Lorentzian and exponential profiles are strongly preferred over the Gaussian model, with Lorentzian wings preferred over exponential wings ($\sim$$3\sigma$). 
    We note, however, that the statistical significance of this preference depends on the specific fitting assumptions, particularly regarding the kinematic coupling between components.
    }
    \label{fig:unite}
\end{figure*}

We model the emission lines using a combination of narrow and broad components. 
All lines are tied to a common redshift and each set of narrow and broad lines share a common FWHM. 
For the permitted transitions (H$\beta$, H$\alpha$, Pa$\gamma$, Pa$\beta$, etc.), we test three different line profiles to characterize the broad line region kinematics: a Gaussian profile, a Lorentzian profile, and a exponential profile.
As shown in Figure~\ref{fig:unite} and Table~\ref{tab:lines}, the non-Gaussian profiles provide a significantly better fit, with the Lorentzian profile preferred at the $2.7\sigma$ level based on the relative Widely Applicable Information Criteria \citep[WAICs;][]{watanabe_AsymptoticEquivalenceBayes_2010}, suggesting that the lines show significant non-Gaussian wings. 
However, we note that the significance of this preference depends on the exact fitting prescription used, such as the choice of priors and the relative independence of the broad and narrow component kinematics.
Follow-up grating spectroscopy at high S/N would be necessary to differentiate between Lorentzian versus exponential wings and constrain the true shape of the broad components in the emission lines. 

\begin{table}[ht!]
\caption{\texttt{unite} Emission Line Fitting Results}
\begin{tabular}{lcccccc}
\hline
\hline
 & \multicolumn{2}{c}{Broad Lorentzian} & \multicolumn{2}{c}{Broad Exponential} & \multicolumn{2}{c}{Broad Gaussian} \\
\hline
WAIC & \multicolumn{2}{c}{30616} & \multicolumn{2}{c}{30626} & \multicolumn{2}{c}{30812} \\
Redshift & \multicolumn{2}{c}{$3.278110^{+0.000095}_{-0.000098}$} & \multicolumn{2}{c}{$3.278114^{+0.000084}_{-0.000098}$} & \multicolumn{2}{c}{$3.278121^{+0.000099}_{-0.000097}$} \\
Narrow FWHM [km\,s$^{-1}$] & \multicolumn{2}{c}{$425^{+17}_{-15}$} & \multicolumn{2}{c}{$430^{+16}_{-17}$} & \multicolumn{2}{c}{$454^{+16}_{-17}$} \\
Broad FWHM [km\,s$^{-1}$] & \multicolumn{2}{c}{$3222^{+82}_{-78}$} & \multicolumn{2}{c}{$2712^{+57}_{-59}$} & \multicolumn{2}{c}{$5070^{+110}_{-100}$} \\
\hline
&\multicolumn{6}{c}{Narrow Line Fluxes [10$^{-20}$ erg\,s$^{-1}$\,cm$^{-2}$] and EWs [\AA]} \\
 & Flux & EW & Flux & EW & Flux & EW \\
\hline
C\,{\sc iv}\,$\lambda$1549\AA & $280^{+120}_{-130}$ & $25^{+14}_{-12}$ & $290^{+120}_{-120}$ & $26^{+13}_{-12}$ & $280^{+120}_{-120}$ & $26^{+14}_{-12}$ \\
C\,{\sc iii}\,$\lambda$1909\AA & $412^{+68}_{-65}$ & $43.2^{+8.3}_{-7.7}$ & $406^{+62}_{-64}$ & $42.6^{+7.3}_{-7.6}$ & $410^{+66}_{-64}$ & $43.1^{+8.4}_{-7.9}$ \\
He\,{\sc ii}\,$\lambda$4687\AA & $92^{+36}_{-35}$ & $3.6^{+1.4}_{-1.3}$ & $101^{+39}_{-40}$ & $3.9^{+1.5}_{-1.5}$ & $108^{+37}_{-41}$ & $4.2^{+1.5}_{-1.6}$ \\
H$\beta$ & $104^{+56}_{-51}$ & $4.0^{+2.1}_{-2.0}$ & $93^{+56}_{-49}$ & $3.6^{+2.2}_{-1.9}$ & $152^{+53}_{-54}$ & $5.9^{+2.0}_{-2.1}$ \\
{[O\,{\sc iii}]}\,$\lambda$4960\AA & $727^{+16}_{-15}$ & $28.08^{+0.66}_{-0.64}$ & $725^{+16}_{-15}$ & $27.84^{+0.65}_{-0.65}$ & $736^{+15}_{-16}$ & $28.27^{+0.65}_{-0.70}$ \\
{[O\,{\sc iii}]}\,$\lambda$5008\AA & $2165^{+47}_{-45}$ & $83.6^{+2.0}_{-1.9}$ & $2160^{+46}_{-46}$ & $82.9^{+1.9}_{-2.0}$ & $2194^{+45}_{-47}$ & $84.2^{+1.9}_{-2.1}$ \\
{[O\,{\sc i}]}\,$\lambda$6302\AA & $33^{+30}_{-21}$ & $1.34^{+1.24}_{-0.84}$ & $87^{+34}_{-32}$ & $3.4^{+1.3}_{-1.3}$ & $95^{+41}_{-38}$ & $3.7^{+1.6}_{-1.5}$ \\
H$\alpha$ & $387^{+98}_{-103}$ & $16.0^{+4.0}_{-4.3}$ & $215^{+96}_{-94}$ & $8.7^{+3.9}_{-3.8}$ & $992^{+95}_{-94}$ & $39.7^{+3.9}_{-3.8}$ \\
He\,{\sc i}\,$\lambda$7067\AA & $16^{+20}_{-12}$ & $0.70^{+0.88}_{-0.51}$ & $15^{+20}_{-11}$ & $0.62^{+0.86}_{-0.47}$ & $20^{+26}_{-15}$ & $0.86^{+1.08}_{-0.62}$ \\
O\,{\sc i}\,$\lambda$8446\AA & $34^{+24}_{-22}$ & $1.8^{+1.2}_{-1.1}$ & $38^{+26}_{-21}$ & $1.9^{+1.3}_{-1.1}$ & $54^{+25}_{-22}$ & $2.7^{+1.2}_{-1.1}$ \\
Pa$\zeta$ & $23^{+31}_{-17}$ & $1.29^{+1.76}_{-0.93}$ & $25^{+32}_{-19}$ & $1.4^{+1.7}_{-1.1}$ & $27^{+28}_{-20}$ & $1.5^{+1.5}_{-1.1}$ \\
Pa$\epsilon$ & $48^{+29}_{-26}$ & $2.8^{+1.7}_{-1.5}$ & $52^{+33}_{-28}$ & $3.0^{+1.9}_{-1.6}$ & $64^{+28}_{-27}$ & $3.7^{+1.6}_{-1.5}$ \\
Pa$\delta$ & $19^{+21}_{-13}$ & $1.20^{+1.30}_{-0.82}$ & $16^{+20}_{-11}$ & $1.01^{+1.27}_{-0.69}$ & $27^{+24}_{-18}$ & $1.7^{+1.5}_{-1.1}$ \\
He\,{\sc i}\,$\lambda$10830\AA & $281^{+43}_{-37}$ & $20.1^{+3.1}_{-2.6}$ & $249^{+39}_{-40}$ & $17.5^{+2.8}_{-2.8}$ & $396^{+40}_{-42}$ & $27.8^{+2.8}_{-2.9}$ \\
Pa$\gamma$ & $37^{+34}_{-26}$ & $2.7^{+2.5}_{-1.9}$ & $25^{+28}_{-18}$ & $1.8^{+2.0}_{-1.3}$ & $29^{+27}_{-20}$ & $2.1^{+1.9}_{-1.5}$ \\
O\,{\sc i}\,$\lambda$11287\AA & $43^{+26}_{-25}$ & $3.3^{+2.0}_{-1.9}$ & $52^{+24}_{-25}$ & $4.0^{+1.9}_{-1.9}$ & $53^{+26}_{-25}$ & $4.1^{+2.0}_{-1.9}$ \\
Pa$\beta$ & $70^{+96}_{-52}$ & $7.5^{+10.3}_{-5.5}$ & $71^{+100}_{-53}$ & $7.6^{+10.7}_{-5.6}$ & $123^{+147}_{-90}$ & $13.2^{+15.8}_{-9.7}$ \\
\hline
&\multicolumn{6}{c}{Broad Line Fluxes [10$^{-20}$ erg\,s$^{-1}$\,cm$^{-2}$] and EWs [\AA]} \\
\hline
H$\beta$ & $1490^{+110}_{-140}$ & $57.6^{+4.5}_{-5.3}$ & $1290^{+110}_{-100}$ & $49.7^{+4.2}_{-4.2}$ & $1170^{+110}_{-100}$ & $44.8^{+4.4}_{-3.9}$ \\
H$\alpha$ & $14930^{+200}_{-210}$ & $618.7^{+9.6}_{-9.9}$ & $13210^{+150}_{-190}$ & $534.5^{+7.7}_{-8.4}$ & $11530^{+160}_{-170}$ & $461.5^{+7.3}_{-6.7}$ \\
He\,{\sc i}\,$\lambda$7067\AA & $465^{+80}_{-87}$ & $20.3^{+3.6}_{-3.8}$ & $448^{+75}_{-73}$ & $19.1^{+3.2}_{-3.1}$ & $386^{+70}_{-70}$ & $16.3^{+3.0}_{-3.0}$ \\
O\,{\sc i}\,$\lambda$8446\AA & $426^{+78}_{-82}$ & $21.7^{+4.0}_{-4.2}$ & $292^{+75}_{-72}$ & $14.5^{+3.8}_{-3.6}$ & $239^{+74}_{-68}$ & $11.8^{+3.7}_{-3.3}$ \\
Pa$\zeta$ & $80^{+86}_{-85}$ & $4.5^{+4.8}_{-4.8}$ & $13^{+75}_{-85}$ & $0.7^{+4.2}_{-4.7}$ & $-22^{+74}_{-68}$ & $-1.2^{+4.0}_{-3.7}$ \\
Pa$\epsilon$ & $163^{+88}_{-96}$ & $9.6^{+5.2}_{-5.6}$ & $87^{+81}_{-81}$ & $5.0^{+4.7}_{-4.7}$ & $43^{+73}_{-78}$ & $2.5^{+4.2}_{-4.5}$ \\
Pa$\delta$ & $486^{+82}_{-77}$ & $30.7^{+5.2}_{-4.9}$ & $406^{+69}_{-72}$ & $25.2^{+4.3}_{-4.5}$ & $378^{+64}_{-71}$ & $23.3^{+3.9}_{-4.4}$ \\
He\,{\sc i}\,$\lambda$10830\AA & $2500^{+140}_{-140}$ & $179^{+10}_{-10}$ & $2230^{+120}_{-110}$ & $157.1^{+8.6}_{-8.0}$ & $1900^{+110}_{-120}$ & $133.2^{+7.4}_{-8.5}$ \\
Pa$\gamma$ & $810^{+110}_{-130}$ & $58.8^{+8.5}_{-9.7}$ & $689^{+98}_{-109}$ & $49.5^{+6.9}_{-7.9}$ & $714^{+96}_{-98}$ & $51.0^{+6.9}_{-7.1}$ \\
Pa$\beta$ & $3670^{+500}_{-480}$ & $396^{+57}_{-53}$ & $3120^{+390}_{-450}$ & $336^{+46}_{-48}$ & $2750^{+380}_{-370}$ & $295^{+45}_{-40}$ \\
\hline
\hline
\end{tabular}
\label{tab:lines}
\end{table}

\section{SED Fitting} \label{app:sed}

This appendix provides additional details of the SED fitting discussed in \S~\ref{ssec:sed}, including the the configuration choices for both CIGALE and AGNFitter.
As emphasized in \S~\ref{ssec:sed}, these fits are intentionally performed on the broadband photometry and X-ray constraints.

\subsection{CIGALE configuration}

We run CIGALE in \texttt{pdf\_analysis} mode with the configuration summarized in Table~\ref{tab:cigale}.
In brief, we adopt a delayed-$\tau$ star-formation history (\texttt{sfhdelayed}), \citet{bruzual_StellarPopulationSynthesis_2003} stellar population synthesis (\texttt{bc03} with a \citet{chabrier_GalacticStellarSubstellar_2003} IMF), a modified starburst attenuation curve (\texttt{dustatt\_modified\_starburst}), THEMIS dust emission (\texttt{themis}; \citealt{jones_GlobalDustModelling_2017}), and SKIRTOR torus models (\texttt{skirtor2016}; \citealt{stalevski_3DRadiativeTransfer_2012,stalevski_DustCoveringFactor_2016}).
We incorporate X-ray constraints using the \citet{lopez_CIGALEModuleTailored_2024} module (\texttt{lopez24}), which ties $L_{\rm 2-10\,keV}$ to the mid-IR luminosity through an $\alpha_{\rm irx}$ parameter and generates an intrinsic AGN X-ray spectrum.
Upper limits are included using CIGALE's upper-limit treatment (we use the \texttt{full} option).
Reported parameter values in Table~\ref{tab:cigale} correspond to the CIGALE Bayesian estimates.

\begin{table}[ht!]
\caption{CIGALE Configuration and Results}
\begin{center}
\begin{tabular}{lccc}
\hline
\hline
\multicolumn{1}{c}{Parameter Name} & Unit & Grid / Option & Measured Value \\
\hline
\multicolumn{4}{c}{Star Formation History: Delayed Star Formation} \\
Main $e$-fold Time ($\tau_{\rm main}$) & Myr & 3000, 5000, 7000, 9000, 11000 & $6000 \pm 2800$ \\
Main Age (age$_{\rm main}$) & Myr & 100, 500, 1000, 3000 & $1000.0 \pm 0.25$ \\
Burst Fraction ($f_{\rm burst}$) & --- & 0.0 & --- \\
Normalize & --- & True & --- \\
Stellar Mass ($M_\star$) & $\rm M_\odot$ & --- & $(2.48 \pm 0.12)\times10^{10}$ \\
\hline 
\multicolumn{4}{c}{Simple Stellar Population: \citet{bruzual_StellarPopulationSynthesis_2003}} \\
Initial Mass Function & --- & 1 \citep{chabrier_GalacticStellarSubstellar_2003} & --- \\ 
Metallicity & $Z$ & 0.008, 0.02 & $0.00800 \pm 0.00021$ \\
Separation Age & Myr & 10 & --- \\
\hline
\multicolumn{4}{c}{Dust Attenuation: Modified \citet{calzetti_DustContentOpacity_2000}} \\
$\rm E(B-V)_{\rm lines}$ & mag & 0.1, 0.3, 0.5, 0.7, 0.9 & $0.7$ \\
Extinction Factor $\left(f_{\rm E(B-V)}\right)$ & --- & 0.1, 0.44, 0.7 & $0.7$ \\
UV Bump Wavelength & nm & 217.5 & --- \\
UV Bump Width & nm & 35.0 & --- \\
UV Bump Amplitude & --- & 0.0 & --- \\
Power-law Slope & --- & $-0.4$, 0.0 & $-0.4$ \\
\hline
\multicolumn{4}{c}{Dust Emission: THEMIS \citep{jones_GlobalDustModelling_2017}} \\
Hydrocarnbon Mass Fraction ($q_{\rm hac}$) & --- & 0.02, 0.17, 0.24 & $0.02000 \pm 0.00014$ \\
Minimum Radiation Field ($U_{\rm min}$) & --- & 0.1, 1.0, 10.0, 50.0 & $30 \pm 20$ \\
Power-law Slope ($\alpha$) & --- & 2.1, 2.5, 2.9 & $2.67 \pm 0.24$ \\
Illumination Fraction ($\gamma$) & --- & 0.4 & --- \\
\hline
\multicolumn{4}{c}{AGN Structure: SKIRTOR \citep{stalevski_3DRadiativeTransfer_2012, stalevski_DustCoveringFactor_2016}} \\
Optical Depth ($\rm\tau_{9.7\mu m}$) & --- & 3, 7, 11 & $8.0\pm3.0$ \\
Radial Gradient ($p$) & --- & 1.0 & --- \\
Polar Gradient ($q$) & --- & 1.0 & --- \\
Opening Angle ($\theta_{\rm open}$) & deg & 20, 40, 60, 80 & $70\pm14$ \\
Radius Ratio ($R_{\rm out}/R_{\rm in}$) & --- & 10, 20, 30 & $18.0\pm7.5$ \\
Clump Mass Fraction ($M_{\rm cl}$) & --- & 0.97 & --- \\
Inclination ($i$) & deg & 10, 30, 50, 70, 90 & $73\pm20$ \\
Disk Type & --- & 2 \citep{lopez_CIGALEModuleTailored_2024} & --- \\
Disk Tuning Parameter ($\delta$) & --- & 0, 0.25, 0.5 & $0.24 \pm 0.21$ \\
AGN Fraction & --- & 0.1, 0.3, 0.5, 0.7, 0.9 & $0.1002  \pm 0.0061$ \\
Polar $\rm E(B-V)$ & mag & 0.00 & --- \\
\hline
\multicolumn{4}{c}{X-ray: \citet{lopez_CIGALEModuleTailored_2024}} \\
Photon Index ($\Gamma$) & --- & 1.8 & --- \\
Cutoff Energy ($E_{\rm cut}$) & keV & 300 & --- \\
$\alpha_{\rm irx}$ Deviation & dex & 0, 0.15, 0.3, 0.45, 0.6 & $0.017 \pm 0.050$ \\
High-mass X-ray Binary Deviation & dex & 0.0 & --- \\
High-mass X-ray Binary Deviation  & dex & 0.0 & --- \\
\hline
\hline
\end{tabular}
\label{tab:cigale}
\end{center}
\end{table}

\subsection{AGNFitter configuration}

We fit the same photometric data with AGNFitter \citep{calistrorivera_AGNfitterBayesianMCMC_2016} using the radio-to-X-ray extension described in \citet{martinez-ramirez_AGNFITTERRXModelingRadiotoXray_2024}.
The model decomposes the SED into four components: a stellar population, cold dust associated with star formation, an accretion disk (``big blue bump''), and a dusty torus.
For the component templates we use \citet{bruzual_StellarPopulationSynthesis_2003} stellar populations, \citet{schreiber_EGGHatchingMock_2017} star-forming dust templates, the accretion disk models of \citet{temple_ModellingType1_2021}, and SKIRTOR torus models.
The implied AGN X-ray luminosity is linked to the mid-IR (6\,$\mu$m) luminosity using the \citet{stern_XRayMidinfraredRelation_2015} relation, enabling a direct comparison between the XRD's observed $L_{\rm 2-10\,keV}$ and the model-predicted value.
Posterior summaries for key parameters are given in Table~\ref{tab:agnfitter}.

\begin{table}[ht!]
\caption{AGNFitter Configuration and Results}
\begin{center}
\begin{tabular}{lcc}
\hline
\hline
\multicolumn{1}{c}{Parameter Name} & Unit & Measured Value \\
\hline
\multicolumn{3}{c}{Galaxy: \citet{bruzual_StellarPopulationSynthesis_2003}} \\
Metallicity & $Z$ &  $0.77^{+0.68}_{-0.48}$ \\ 
$e$-fold Time ($\tau$) & --- & $4.0^{+3.3}_{-2.5}$ \\
Age & Myr & $720^{+760}_{-610}$ \\ 
$\rm E(B-V)_{\rm gal}$ & mag & $0.153^{+0.097}_{-0.070}$ \\
Stellar Mass ($M_\star$) & $\rm M_\odot$ & $\left(1.4^{+2.6}_{-1.2}\right) \times 10^9$\\
\hline
\multicolumn{3}{c}{Starburst: \citet{schreiber_EGGHatchingMock_2017}} \\
Dust Temperature & K & $31.2^{+7.4}_{-10.1}$ \\ 
PAH Fraction ($f_{\rm PAH}$) & --- & $0.027^{+0.018}_{-0.018}$ \\
\hline
\multicolumn{3}{c}{Big Blue Bump: \citet{temple_ModellingType1_2021}} \\
$\rm E(B-V)_{\rm BBB}$ & mag & $0.885^{+0.018}_{-0.020}$ \\
\hline
\multicolumn{3}{c}{Torus: SKIRTOR \citep{stalevski_3DRadiativeTransfer_2012, stalevski_DustCoveringFactor_2016}} \\
Optical Depth ($\rm\tau_{v}$) & --- & $10.47^{+0.35}_{-0.34}$ \\
Opening Angle ($oa$) & deg & $12.66^{+1.69}_{-1.79}$ \\
Inclination ($i$) & deg & $30.14^{+3.53}_{-3.50}$ \\
\hline
\hline
\end{tabular}
\label{tab:agnfitter}
\end{center}
\end{table}

\bibliography{bibliography}{}

\begin{thebibliography}{}
\expandafter\ifx\csname natexlab\endcsname\relax\def\natexlab#1{#1}\fi
\providecommand{\url}[1]{\href{#1}{#1}}
\providecommand{\dodoi}[1]{doi:~\href{http://doi.org/#1}{\nolinkurl{#1}}}
\providecommand{\doeprint}[1]{\href{http://ascl.net/#1}{\nolinkurl{http://ascl.net/#1}}}
\providecommand{\doarXiv}[1]{\href{https://arxiv.org/abs/#1}{\nolinkurl{https://arxiv.org/abs/#1}}}

\bibitem[{H.~B. Akins {et~al.}(2024)Akins, Casey, Berg, Chisholm, Franco, Finkelstein, Fujimoto, Kokorev, Lambrides, Robertson, Taylor, Coulter, Fox, \& Karmen}]{akins_StrongRestUVEmission_2024}
Akins, H.~B., Casey, C.~M., Berg, D.~A., {et~al.} 2024, \bibinfo{title}{Strong rest-{UV} emission lines in a “little red dot” {AGN} at {\textless}span class="nocase"{\textgreater}z=7{\textless}/span{\textgreater}: {Early} {SMBH} growth alongside compact massive star formation?} arXiv e-prints, arXiv:2410.00949, \dodoi{10.48550/arXiv.2410.00949}

\bibitem[{H.~B. Akins {et~al.}(2025)Akins, Casey, Lambrides, Allen, Andika, Brinch, Champagne, Cooper, Ding, Drakos, Faisst, Finkelstein, Franco, Fujimoto, Gentile, Gillman, Gozaliasl, Harish, Hayward, Hirschmann, Ilbert, Kartaltepe, Kocevski, Koekemoer, Kokorev, Liu, Long, McCracken, McKinney, Onoue, Paquereau, Renzini, Rhodes, Robertson, Shuntov, Silverman, Tanaka, Toft, Trakhtenbrot, Valentino, \& Zavala}]{akins_COSMOSWebOverabundancePhysical_2025}
Akins, H.~B., Casey, C.~M., Lambrides, E., {et~al.} 2025, \bibinfo{title}{{COSMOS}-{Web}: {The} {Overabundance} and {Physical} {Nature} of "{Little} {Red} {Dots}"—{Implications} for {Early} {Galaxy} and {SMBH} {Assembly},} The Astrophysical Journal, 991, 37, \dodoi{10.3847/1538-4357/ade984}

\bibitem[{T.~T. Ananna {et~al.}(2024)Ananna, Bogdán, Kovács, Natarajan, \& Hickox}]{ananna_XRayViewLittle_2024}
Ananna, T.~T., Bogdán, A., Kovács, O.~E., Natarajan, P., \& Hickox, R.~C. 2024, \bibinfo{title}{X-{Ray} {View} of {Little} {Red} {Dots}: {Do} {They} {Host} {Supermassive} {Black} {Holes}?} The Astrophysical Journal, 969, L18, \dodoi{10.3847/2041-8213/ad5669}

\bibitem[{K.~A. Arnaud(1996)Arnaud}]{arnaud_XSPECFirstTen_1996}
Arnaud, K.~A. 1996, \bibinfo{title}{{XSPEC}: {The} {First} {Ten} {Years},} in Astronomical {Data} {Analysis} {Software} and {Systems} {V}, Vol. 101, 17

\bibitem[{ {Astropy Collaboration} {et~al.}(2013){Astropy Collaboration}, Robitaille, Tollerud, Greenfield, Droettboom, Bray, Aldcroft, Davis, Ginsburg, Price-Whelan, Kerzendorf, Conley, Crighton, Barbary, Muna, Ferguson, Grollier, Parikh, Nair, Unther, Deil, Woillez, Conseil, Kramer, Turner, Singer, Fox, Weaver, Zabalza, Edwards, Azalee~Bostroem, Burke, Casey, Crawford, Dencheva, Ely, Jenness, Labrie, Lim, Pierfederici, Pontzen, Ptak, Refsdal, Servillat, \& Streicher}]{astropycollaboration_AstropyCommunityPython_2013}
{Astropy Collaboration}, Robitaille, T.~P., Tollerud, E.~J., {et~al.} 2013, \bibinfo{title}{Astropy: {A} community {Python} package for astronomy,} Astronomy and Astrophysics, 558, A33, \dodoi{10.1051/0004-6361/201322068}

\bibitem[{ {Astropy Collaboration} {et~al.}(2018){Astropy Collaboration}, Price-Whelan, Sipőcz, Günther, Lim, Crawford, Conseil, Shupe, Craig, Dencheva, Ginsburg, VanderPlas, Bradley, Pérez-Suárez, de~Val-Borro, Aldcroft, Cruz, Robitaille, Tollerud, Ardelean, Babej, Bach, Bachetti, Bakanov, Bamford, Barentsen, Barmby, Baumbach, Berry, Biscani, Boquien, Bostroem, Bouma, Brammer, Bray, Breytenbach, Buddelmeijer, Burke, Calderone, Cano~Rodríguez, Cara, Cardoso, Cheedella, Copin, Corrales, Crichton, D'Avella, Deil, Depagne, Dietrich, Donath, Droettboom, Earl, Erben, Fabbro, Ferreira, Finethy, Fox, Garrison, Gibbons, Goldstein, Gommers, Greco, Greenfield, Groener, Grollier, Hagen, Hirst, Homeier, Horton, Hosseinzadeh, Hu, Hunkeler, Ivezić, Jain, Jenness, Kanarek, Kendrew, Kern, Kerzendorf, Khvalko, King, Kirkby, Kulkarni, Kumar, Lee, Lenz, Littlefair, Ma, Macleod, Mastropietro, McCully, Montagnac, Morris, Mueller, Mumford, Muna, Murphy, Nelson, Nguyen, Ninan, Nöthe, Ogaz, Oh, Parejko, Parley, Pascual, Patil, Patil, Plunkett, Prochaska, Rastogi, Reddy~Janga, Sabater, Sakurikar, Seifert, Sherbert, Sherwood-Taylor, Shih, Sick, Silbiger, Singanamalla, Singer, Sladen, Sooley, Sornarajah, Streicher, Teuben, Thomas, Tremblay, Turner, Terrón, van Kerkwijk, de~la Vega, Watkins, Weaver, Whitmore, Woillez, Zabalza, \& {Astropy Contributors}}]{astropycollaboration_AstropyProjectBuilding_2018}
{Astropy Collaboration}, Price-Whelan, A.~M., Sipőcz, B.~M., {et~al.} 2018, \bibinfo{title}{The {Astropy} {Project}: {Building} an {Open}-science {Project} and {Status} of the v2.0 {Core} {Package},} The Astronomical Journal, 156, 123, \dodoi{10.3847/1538-3881/aabc4f}

\bibitem[{ {Astropy Collaboration} {et~al.}(2022){Astropy Collaboration}, Price-Whelan, Lim, Earl, Starkman, Bradley, Shupe, Patil, Corrales, Brasseur, Nöthe, Donath, Tollerud, Morris, Ginsburg, Vaher, Weaver, Tocknell, Jamieson, van Kerkwijk, Robitaille, Merry, Bachetti, Günther, Aldcroft, Alvarado-Montes, Archibald, Bódi, Bapat, Barentsen, Bazán, Biswas, Boquien, Burke, Cara, Cara, Conroy, Conseil, Craig, Cross, Cruz, D'Eugenio, Dencheva, Devillepoix, Dietrich, Eigenbrot, Erben, Ferreira, Foreman-Mackey, Fox, Freij, Garg, Geda, Glattly, Gondhalekar, Gordon, Grant, Greenfield, Groener, Guest, Gurovich, Handberg, Hart, Hatfield-Dodds, Homeier, Hosseinzadeh, Jenness, Jones, Joseph, Kalmbach, Karamehmetoglu, Kałuszyński, Kelley, Kern, Kerzendorf, Koch, Kulumani, Lee, Ly, Ma, MacBride, Maljaars, Muna, Murphy, Norman, O'Steen, Oman, Pacifici, Pascual, Pascual-Granado, Patil, Perren, Pickering, Rastogi, Roulston, Ryan, Rykoff, Sabater, Sakurikar, Salgado, Sanghi, Saunders, Savchenko, Schwardt, Seifert-Eckert, Shih, Jain, Shukla, Sick, Simpson, Singanamalla, Singer, Singhal, Sinha, Sipőcz, Spitler, Stansby, Streicher, Šumak, Swinbank, Taranu, Tewary, Tremblay, de~Val-Borro, Van~Kooten, Vasović, Verma, de~Miranda~Cardoso, Williams, Wilson, Winkel, Wood-Vasey, Xue, Yoachim, Zhang, Zonca, \& {Astropy Project Contributors}}]{astropycollaboration_AstropyProjectSustaining_2022}
{Astropy Collaboration}, Price-Whelan, A.~M., Lim, P.~L., {et~al.} 2022, \bibinfo{title}{The {Astropy} {Project}: {Sustaining} and {Growing} a {Community}-oriented {Open}-source {Project} and the {Latest} {Major} {Release} (v5.0) of the {Core} {Package},} The Astrophysical Journal, 935, 167, \dodoi{10.3847/1538-4357/ac7c74}

\bibitem[{ {Astropy-Specutils Development Team}(2019){Astropy-Specutils Development Team}}]{astropy-specutilsdevelopmentteam_SpecutilsSpectroscopicAnalysis_2019}
{Astropy-Specutils Development Team}. 2019, \bibinfo{title}{Specutils: {Spectroscopic} analysis and reduction,} Astrophysics Source Code Library, ascl:1902.012

\bibitem[{M.~C. Begelman \& J. Dexter(2026)Begelman \& Dexter}]{begelman_LittleRedDots_2026}
Begelman, M.~C., \& Dexter, J. 2026, \bibinfo{title}{Little {Red} {Dots} as {Late}-stage {Quasi}-stars,} The Astrophysical Journal, 996, 48, \dodoi{10.3847/1538-4357/ae274a}

\bibitem[{M. Boquien {et~al.}(2019)Boquien, Burgarella, Roehlly, Buat, Ciesla, Corre, Inoue, \& Salas}]{boquien_CIGALEPythonCode_2019}
Boquien, M., Burgarella, D., Roehlly, Y., {et~al.} 2019, \bibinfo{title}{{CIGALE}: a python {Code} {Investigating} {GALaxy} {Emission},} Astronomy and Astrophysics, 622, A103, \dodoi{10.1051/0004-6361/201834156}

\bibitem[{J. Bradbury {et~al.}(2018)Bradbury, Frostig, Hawkins, Johnson, Leary, Maclaurin, Necula, Paszke, VanderPlas, Wanderman-Milne, \& Zhang}]{bradbury_JAXComposableTransformations_2018}
Bradbury, J., Frostig, R., Hawkins, P., {et~al.} 2018, {JAX}: {Composable} transformations of {Python}+{NumPy} programs,

\bibitem[{G. Brammer(2023)Brammer}]{brammer_MsaexpNIRSpecAnalyis_2023}
Brammer, G. 2023, Msaexp: {NIRSpec} analyis tools, \dodoi{10.5281/zenodo.8319596}

\bibitem[{G. Brammer \& F. Valentino(2025)Brammer \& Valentino}]{brammer_DAWNJWSTArchive_2025}
Brammer, G., \& Valentino, F. 2025, The {DAWN} {JWST} {Archive}: {Compilation} of {Public} {NIRSpec} {Spectra}, Zenodo, \dodoi{10.5281/zenodo.15472354}

\bibitem[{G.~B. Brammer {et~al.}(2012)Brammer, van Dokkum, Franx, Fumagalli, Patel, Rix, Skelton, Kriek, Nelson, Schmidt, Bezanson, da~Cunha, Erb, Fan, Förster~Schreiber, Illingworth, Labbé, Leja, Lundgren, Magee, Marchesini, McCarthy, Momcheva, Muzzin, Quadri, Steidel, Tal, Wake, Whitaker, \& Williams}]{brammer_3DHSTWidefieldGrism_2012}
Brammer, G.~B., van Dokkum, P.~G., Franx, M., {et~al.} 2012, \bibinfo{title}{{3D}-{HST}: {A} {Wide}-field {Grism} {Spectroscopic} {Survey} with the {Hubble} {Space} {Telescope},} The Astrophysical Journal Supplement Series, 200, 13, \dodoi{10.1088/0067-0049/200/2/13}

\bibitem[{W.~N. Brandt \& D.~M. Alexander(2015)Brandt \& Alexander}]{brandt_CosmicXraySurveys_2015}
Brandt, W.~N., \& Alexander, D.~M. 2015, \bibinfo{title}{Cosmic {X}-ray surveys of distant active galaxies. {The} demographics, physics, and ecology of growing supermassive black holes,} Astronomy and Astrophysics Review, 23, 1, \dodoi{10.1007/s00159-014-0081-z}

\bibitem[{G. Bruzual \& S. Charlot(2003)Bruzual \& Charlot}]{bruzual_StellarPopulationSynthesis_2003}
Bruzual, G., \& Charlot, S. 2003, \bibinfo{title}{Stellar population synthesis at the resolution of 2003,} Monthly Notices RAS, 344, 1000, \dodoi{10.1046/j.1365-8711.2003.06897.x}

\bibitem[{D. Burgarella {et~al.}(2005)Burgarella, Buat, \& Iglesias-Páramo}]{burgarella_StarFormationDust_2005}
Burgarella, D., Buat, V., \& Iglesias-Páramo, J. 2005, \bibinfo{title}{Star formation and dust attenuation properties in galaxies from a statistical ultraviolet-to-far-infrared analysis,} Monthly Notices of the Royal Astronomical Society, 360, 1413, \dodoi{10.1111/j.1365-2966.2005.09131.x}

\bibitem[{G. Calistro~Rivera {et~al.}(2016)Calistro~Rivera, Lusso, Hennawi, \& Hogg}]{calistrorivera_AGNfitterBayesianMCMC_2016}
Calistro~Rivera, G., Lusso, E., Hennawi, J.~F., \& Hogg, D.~W. 2016, \bibinfo{title}{{AGNfitter}: {A} {Bayesian} {MCMC} {Approach} to {Fitting} {Spectral} {Energy} {Distributions} of {AGNs},} The Astrophysical Journal, 833, 98, \dodoi{10.3847/1538-4357/833/1/98}

\bibitem[{D. Calzetti {et~al.}(2000)Calzetti, Armus, Bohlin, Kinney, Koornneef, \& Storchi-Bergmann}]{calzetti_DustContentOpacity_2000}
Calzetti, D., Armus, L., Bohlin, R.~C., {et~al.} 2000, \bibinfo{title}{The {Dust} {Content} and {Opacity} of {Actively} {Star}-forming {Galaxies},} The Astrophysical Journal, 533, 682, \dodoi{10.1086/308692}

\bibitem[{C.~M. Casey {et~al.}(2024)Casey, Akins, Kokorev, McKinney, Cooper, Long, Franco, \& Manning}]{casey_DustLittleRed_2024}
Casey, C.~M., Akins, H.~B., Kokorev, V., {et~al.} 2024, \bibinfo{title}{Dust in {Little} {Red} {Dots},} The Astrophysical Journal, 975, L4, \dodoi{10.3847/2041-8213/ad7ba7}

\bibitem[{W. Cash(1979)Cash}]{cash_ParameterEstimationAstronomy_1979}
Cash, W. 1979, \bibinfo{title}{Parameter estimation in astronomy through application of the likelihood ratio.,} The Astrophysical Journal, 228, 939, \dodoi{10.1086/156922}

\bibitem[{G. Chabrier(2003)Chabrier}]{chabrier_GalacticStellarSubstellar_2003}
Chabrier, G. 2003, \bibinfo{title}{Galactic {Stellar} and {Substellar} {Initial} {Mass} {Function},} Publications of the Astronomical Society of the Pacific, 115, 763, \dodoi{10.1086/376392}

\bibitem[{S.-J. Chang {et~al.}(2025)Chang, Gronke, Matthee, \& Mason}]{chang_ImpactResonanceRaman_2025}
Chang, S.-J., Gronke, M., Matthee, J., \& Mason, C. 2025, \bibinfo{title}{Impact of {Resonance}, {Raman}, and {Thomson} {Scattering} on {Hydrogen} {Line} {Formation} in {Little} {Red} {Dots},} arXiv, arXiv:2508.08768, \dodoi{10.48550/arXiv.2508.08768}

\bibitem[{J. Chevallard {et~al.}(2013)Chevallard, Charlot, Wandelt, \& Wild}]{chevallard_InsightsContentSpatial_2013}
Chevallard, J., Charlot, S., Wandelt, B., \& Wild, V. 2013, \bibinfo{title}{Insights into the content and spatial distribution of dust from the integrated spectral properties of galaxies,} Monthly Notices of the Royal Astronomical Society, 432, 2061, \dodoi{10.1093/mnras/stt523}

\bibitem[{M. Chira {et~al.}(2026)Chira, Georgakakis, Ruiz, Chen, Buchner, Rankine, Kammoun, Aydar, Salvato, Merloni, \& Krumpe}]{chira_RevisitingXraytoUVRelation_2026}
Chira, M., Georgakakis, A., Ruiz, A., {et~al.} 2026, \bibinfo{title}{Revisiting the {X}-ray-to-{UV} relation of quasars in the era of all-sky surveys,} Monthly notices of the Royal Astronomical Society, 545, staf1905, \dodoi{10.1093/mnras/staf1905}

\bibitem[{B. Czerny {et~al.}(2004)Czerny, Li, Loska, \& Szczerba}]{czerny_ExtinctionDueAmorphous_2004}
Czerny, B., Li, J., Loska, Z., \& Szczerba, R. 2004, \bibinfo{title}{Extinction due to amorphous carbon grains in red quasars from the {Sloan} {Digital} {Sky} {Survey},} Monthly Notices of the Royal Astronomical Society, 348, L54, \dodoi{10.1111/j.1365-2966.2004.07590.x}

\bibitem[{M. Davis {et~al.}(2007)Davis, Guhathakurta, Konidaris, Newman, Ashby, Biggs, Barmby, Bundy, Chapman, Coil, Conselice, Cooper, Croton, Eisenhardt, Ellis, Faber, Fang, Fazio, Georgakakis, Gerke, Goss, Gwyn, Harker, Hopkins, Huang, Ivison, Kassin, Kirby, Koekemoer, Koo, Laird, Le~Floc'h, Lin, Lotz, Marshall, Martin, Metevier, Moustakas, Nandra, Noeske, Papovich, Phillips, Rich, Rieke, Rigopoulou, Salim, Schiminovich, Simard, Smail, Small, Weiner, Willmer, Willner, Wilson, Wright, \& Yan}]{davis_AllWavelengthExtendedGroth_2007}
Davis, M., Guhathakurta, P., Konidaris, N.~P., {et~al.} 2007, \bibinfo{title}{The {All}-{Wavelength} {Extended} {Groth} {Strip} {International} {Survey} ({AEGIS}) {Data} {Sets},} The Astrophysical Journal, 660, L1, \dodoi{10.1086/517931}

\bibitem[{A. de~Graaff {et~al.}(2024)de~Graaff, Rix, Carniani, Suess, Charlot, Curtis-Lake, Arribas, Baker, Boyett, Bunker, Cameron, Chevallard, Curti, Eisenstein, Franx, Hainline, Hausen, Ji, Johnson, Jones, Maiolino, Maseda, Nelson, Parlanti, Rawle, Robertson, Tacchella, Übler, Williams, Willmer, \& Willott}]{degraaff_IonisedGasKinematics_2024}
de~Graaff, A., Rix, H.-W., Carniani, S., {et~al.} 2024, \bibinfo{title}{Ionised gas kinematics and dynamical masses of z ≳ 6 galaxies from {JADES}/{NIRSpec} high-resolution spectroscopy,} Astronomy and Astrophysics, 684, A87, \dodoi{10.1051/0004-6361/202347755}

\bibitem[{A. de~Graaff {et~al.}(2025{\natexlab{a}})de~Graaff, Hviding, Naidu, Greene, Miller, Leja, Matthee, Brammer, Katz, Bezanson, Boogaard, Bose, Chisholm, Cleri, Dayal, Feldmann, Fudamoto, Fujimoto, Furtak, Glazebrook, Gottumukkala, Heintz, Kokorev, Labbe, Maseda, McConachie, Nanayakkara, Nelson, Nowaczyk, Oesch, Rix, Setton, Torralba, Walter, Wang, Weibel, \& van~der Wel}]{degraaff_LittleRedDots_2025}
de~Graaff, A., Hviding, R.~E., Naidu, R.~P., {et~al.} 2025{\natexlab{a}}, \bibinfo{title}{Little {Red} {Dots} host {Black} {Hole} {Stars}: {A} unified family of gas-reddened {AGN} revealed by {JWST}/{NIRSpec} spectroscopy,} arXiv e-prints, arXiv:2511.21820, \dodoi{10.48550/arXiv.2511.21820}

\bibitem[{A. de~Graaff {et~al.}(2025{\natexlab{b}})de~Graaff, Rix, Naidu, Labbe, Wang, Leja, Matthee, Katz, Greene, Hviding, Baggen, Bezanson, Boogaard, Brammer, Dayal, van Dokkum, Goulding, Hirschmann, Maseda, McConachie, Miller, Nelson, Oesch, Setton, Shivaei, Weibel, Whitaker, \& Williams}]{degraaff_RemarkableRubyAbsorption_2025}
de~Graaff, A., Rix, H.-W., Naidu, R.~P., {et~al.} 2025{\natexlab{b}}, \bibinfo{title}{A remarkable {Ruby}: {Absorption} in dense gas, rather than evolved stars, drives the extreme {Balmer} break of a {Little} {Red} {Dot} at {\textless}span class="nocase"{\textgreater}z=3.5{\textless}/span{\textgreater},} arXiv e-prints, arXiv:2503.16600, \dodoi{10.48550/arXiv.2503.16600}

\bibitem[{A. de~Graaff {et~al.}(2025{\natexlab{c}})de~Graaff, Brammer, Weibel, Lewis, Maseda, Oesch, Bezanson, Boogaard, Cleri, Cooper, Gottumukkala, Greene, Hirschmann, Hviding, Katz, Labbé, Leja, Matthee, McConachie, Miller, Naidu, Price, Rix, Setton, Suess, Wang, Whitaker, \& Williams}]{degraaff_RUBIESCompleteCensus_2025}
de~Graaff, A., Brammer, G., Weibel, A., {et~al.} 2025{\natexlab{c}}, \bibinfo{title}{{RUBIES}: {A} complete census of the bright and red distant {Universe} with {JWST}/{NIRSpec},} Astronomy and Astrophysics, 697, A189, \dodoi{10.1051/0004-6361/202452186}

\bibitem[{F. D'Eugenio {et~al.}(2025)D'Eugenio, Maiolino, Perna, Uebler, Ji, McClymont, Koudmani, Sijacki, Juodžbalis, Scholtz, Bennett, Bunker, Carniani, Charlot, Cresci, Curtis-Lake, Dalla~Bontà, Jones, Lyu, Marconi, Mazzolari, Nelson, Parlanti, Robertson, Schneider, Simmonds, Tacchella, Venturi, Willott, Witstok, \& Witten}]{deugenio_BlackTHUNDERStrikesTwice_2025}
D'Eugenio, F., Maiolino, R., Perna, M., {et~al.} 2025, \bibinfo{title}{{BlackTHUNDER} strikes twice: {Rest}-frame {Balmer}-line absorption and high {Eddington} accretion rate in a {Little} {Red} {Dot} at {\textless}span class="nocase"{\textgreater}z=7.04{\textless}/span{\textgreater},} arXiv e-prints, arXiv:2503.11752, \dodoi{10.48550/arXiv.2503.11752}

\bibitem[{X. Dong {et~al.}(2008)Dong, Wang, Wang, Yuan, Zhou, Dai, \& Zhang}]{dong_BroadlineBalmerDecrements_2008}
Dong, X., Wang, T., Wang, J., {et~al.} 2008, \bibinfo{title}{Broad-line {Balmer} decrements in blue active galactic nuclei,} Monthly Notices of the Royal Astronomical Society, 383, 581, \dodoi{10.1111/j.1365-2966.2007.12560.x}

\bibitem[{F. Duras {et~al.}(2020)Duras, Bongiorno, Ricci, Piconcelli, Shankar, Lusso, Bianchi, Fiore, Maiolino, Marconi, Onori, Sani, Schneider, Vignali, \& La~Franca}]{duras_UniversalBolometricCorrections_2020}
Duras, F., Bongiorno, A., Ricci, F., {et~al.} 2020, \bibinfo{title}{Universal bolometric corrections for active galactic nuclei over seven luminosity decades,} Astronomy and Astrophysics, 636, A73, \dodoi{10.1051/0004-6361/201936817}

\bibitem[{S. Fu {et~al.}(2025)Fu, Zhang, Jiang, Chen, Jiang, Ho, Inayoshi, Chen, Lyu, Sun, Wang, \& Yang}]{fu_DiscoveryTwoLittle_2025}
Fu, S., Zhang, Z., Jiang, D., {et~al.} 2025, Discovery of two little red dots transitioning into quasars, arXiv, \dodoi{10.48550/arXiv.2512.02096}

\bibitem[{P. Gandhi {et~al.}(2009)Gandhi, Horst, Smette, Hönig, Comastri, Gilli, Vignali, \& Duschl}]{gandhi_ResolvingMidinfraredCores_2009}
Gandhi, P., Horst, H., Smette, A., {et~al.} 2009, \bibinfo{title}{Resolving the mid-infrared cores of local {Seyferts},} Astronomy and Astrophysics, 502, 457, \dodoi{10.1051/0004-6361/200811368}

\bibitem[{C.~M. Gaskell {et~al.}(2004)Gaskell, Goosmann, Antonucci, \& Whysong}]{gaskell_NuclearReddeningCurve_2004}
Gaskell, C.~M., Goosmann, R.~W., Antonucci, R. R.~J., \& Whysong, D.~H. 2004, \bibinfo{title}{The {Nuclear} {Reddening} {Curve} for {Active} {Galactic} {Nuclei} and the {Shape} of the {Infrared} to {X}-{Ray} {Spectral} {Energy} {Distribution},} The Astrophysical Journal, 616, 147, \dodoi{10.1086/423885}

\bibitem[{J.~E. Geach {et~al.}(2017)Geach, Dunlop, Halpern, Smail, van~der Werf, Alexander, Almaini, Aretxaga, Arumugam, Asboth, Banerji, Beanlands, Best, Blain, Birkinshaw, Chapin, Chapman, Chen, Chrysostomou, Clarke, Clements, Conselice, Coppin, Cowley, Danielson, Eales, Edge, Farrah, Gibb, Harrison, Hine, Hughes, Ivison, Jarvis, Jenness, Jones, Karim, Koprowski, Knudsen, Lacey, Mackenzie, Marsden, McAlpine, McMahon, Meijerink, Michałowski, Oliver, Page, Peacock, Rigopoulou, Robson, Roseboom, Rotermund, Scott, Serjeant, Simpson, Simpson, Smith, Spaans, Stanley, Stevens, Swinbank, Targett, Thomson, Valiante, Wake, Webb, Willott, Zavala, \& Zemcov}]{geach_SCUBA2CosmologyLegacy_2017}
Geach, J.~E., Dunlop, J.~S., Halpern, M., {et~al.} 2017, \bibinfo{title}{The {SCUBA}-2 {Cosmology} {Legacy} {Survey}: 850 𝜇m maps, catalogues and number counts,} Monthly Notices of the Royal Astronomical Society, 465, 1789, \dodoi{10.1093/mnras/stw2721}

\bibitem[{E. Glikman {et~al.}(2006)Glikman, Helfand, \& White}]{glikman_NearInfraredSpectralTemplate_2006}
Glikman, E., Helfand, D.~J., \& White, R.~L. 2006, \bibinfo{title}{A {Near}-{Infrared} {Spectral} {Template} for {Quasars},} The Astrophysical Journal, 640, 579, \dodoi{10.1086/500098}

\bibitem[{K. Gordon(2024)Gordon}]{gordon_Dust_extinctionInterstellarDust_2024}
Gordon, K. 2024, \bibinfo{title}{Dust\_extinction: {Interstellar} {Dust} {Extinction} {Models},} The Journal of Open Source Software, 9, 7023, \dodoi{10.21105/joss.07023}

\bibitem[{K.~D. Gordon {et~al.}(2003)Gordon, Clayton, Misselt, Landolt, \& Wolff}]{gordon_QuantitativeComparisonSmall_2003}
Gordon, K.~D., Clayton, G.~C., Misselt, K.~A., Landolt, A.~U., \& Wolff, M.~J. 2003, \bibinfo{title}{A {Quantitative} {Comparison} of the {Small} {Magellanic} {Cloud}, {Large} {Magellanic} {Cloud}, and {Milky} {Way} {Ultraviolet} to {Near}-{Infrared} {Extinction} {Curves},} The Astrophysical Journal, 594, 279, \dodoi{10.1086/376774}

\bibitem[{A.~D. Goulding {et~al.}(2012)Goulding, Forman, Hickox, Jones, Kraft, Murray, Vikhlinin, Coil, Cooper, Davis, \& Newman}]{goulding_ChandraXRayPointsource_2012}
Goulding, A.~D., Forman, W.~R., Hickox, R.~C., {et~al.} 2012, \bibinfo{title}{The {Chandra} {X}-{Ray} {Point}-source {Catalog} in the {DEEP2} {Galaxy} {Redshift} {Survey} {Fields},} The Astrophysical Journal Supplement Series, 202, 6, \dodoi{10.1088/0067-0049/202/1/6}

\bibitem[{J.~E. Greene \& L.~C. Ho(2005)Greene \& Ho}]{greene_EstimatingBlackHole_2005}
Greene, J.~E., \& Ho, L.~C. 2005, \bibinfo{title}{Estimating {Black} {Hole} {Masses} in {Active} {Galaxies} {Using} the {Hα} {Emission} {Line},} The Astrophysical Journal, 630, 122, \dodoi{10.1086/431897}

\bibitem[{J.~E. Greene {et~al.}(2025)Greene, Setton, Furtak, Naidu, Volonteri, Dayal, Labbe, van Dokkum, Bezanson, Brammer, \& al.}]{greene_WhatYouSee_2025}
Greene, J.~E., Setton, D.~J., Furtak, L.~J., {et~al.} 2025, \bibinfo{title}{What you see is what you get: empirically measured bolometric luminosities of {Little} {Red} {Dots},} arXiv e-prints, arXiv:2509.05434, \dodoi{10.48550/arXiv.2509.05434}

\bibitem[{C.~M. Gunasekera {et~al.}(2025)Gunasekera, van Hoof, Dehghanian, Chakraborty, Shaw, Bianchi, Chatzikos, Tsujimoto, \& Ferland}]{gunasekera_2025ReleaseCloudy_2025}
Gunasekera, C.~M., van Hoof, P. A.~M., Dehghanian, M., {et~al.} 2025, \bibinfo{title}{The 2025 {Release} of {Cloudy},} arXiv e-prints, arXiv:2508.01102, \dodoi{10.48550/arXiv.2508.01102}

\bibitem[{H. Hao {et~al.}(2010)Hao, Elvis, Civano, Lanzuisi, Brusa, Lusso, Zamorani, Comastri, Bongiorno, Impey, Koekemoer, Le~Floc'h, Salvato, Sanders, Trump, \& Vignali}]{hao_HotdustpoorType1_2010}
Hao, H., Elvis, M., Civano, F., {et~al.} 2010, \bibinfo{title}{Hot-dust-poor {Type} 1 {Active} {Galactic} {Nuclei} in the {COSMOS} {Survey},} The Astrophysical Journal, 724, L59, \dodoi{10.1088/2041-8205/724/1/L59}

\bibitem[{C.~R. Harris {et~al.}(2020)Harris, Millman, van~der Walt, Gommers, Virtanen, Cournapeau, Wieser, Taylor, Berg, Smith, Kern, Picus, Hoyer, van Kerkwijk, Brett, Haldane, del Río, Wiebe, Peterson, Gérard-Marchant, Sheppard, Reddy, Weckesser, Abbasi, Gohlke, \& Oliphant}]{harris_ArrayProgrammingNumPy_2020}
Harris, C.~R., Millman, K.~J., van~der Walt, S.~J., {et~al.} 2020, \bibinfo{title}{Array programming with {NumPy},} Nature, 585, 357, \dodoi{10.1038/s41586-020-2649-2}

\bibitem[{K.~E. Heintz {et~al.}(2025)Heintz, Brammer, Watson, Oesch, Keating, Hayes, {Abdurro'uf}, Arellano-Córdova, Carnall, Christiansen, Cullen, Davé, Dayal, Ferrara, Finlator, Fynbo, Flury, Gelli, Gillman, Gottumukkala, Gould, Greve, Hardin, Hsiao, Hutter, Jakobsson, Killi, Khosravaninezhad, Laursen, Lee, Magdis, Matthee, Naidu, Narayanan, Pollock, Prescott, Rusakov, Shuntov, Sneppen, Smit, Tanvir, Terp, Toft, Valentino, Vijayan, Weaver, Wise, \& Witstok}]{heintz_JWSTPRIMALArchivalSurvey_2025}
Heintz, K.~E., Brammer, G.~B., Watson, D., {et~al.} 2025, \bibinfo{title}{The {JWST}-{PRIMAL} archival survey: {A} {JWST}/{NIRSpec} reference sample for the physical properties and {Lyman}-α absorption and emission of ∼600 galaxies at z = 5.0 - 13.4,} Astronomy and Astrophysics, 693, A60, \dodoi{10.1051/0004-6361/202450243}

\bibitem[{R.~C. Hickox \& D.~M. Alexander(2018)Hickox \& Alexander}]{hickox_ObscuredActiveGalactic_2018}
Hickox, R.~C., \& Alexander, D.~M. 2018, \bibinfo{title}{Obscured {Active} {Galactic} {Nuclei},} Annual Review of Astronomy and Astrophysics, 56, 625, \dodoi{10.1146/annurev-astro-081817-051803}

\bibitem[{M. Hirschmann {et~al.}(2019)Hirschmann, Charlot, Feltre, Naab, Somerville, \& Choi}]{hirschmann_SyntheticNebularEmission_2019}
Hirschmann, M., Charlot, S., Feltre, A., {et~al.} 2019, \bibinfo{title}{Synthetic nebular emission from massive galaxies - {II}. {Ultraviolet}-line diagnostics of dominant ionizing sources,} Monthly Notices of the Royal Astronomical Society, 487, 333, \dodoi{10.1093/mnras/stz1256}

\bibitem[{K. Horne(1986)Horne}]{horne_OptimalExtractionAlgorithm_1986}
Horne, K. 1986, \bibinfo{title}{An optimal extraction algorithm for {CCD} spectroscopy.,} Publications of the Astronomical Society of the Pacific, 98, 609, \dodoi{10.1086/131801}

\bibitem[{J.~D. Hunter(2007)Hunter}]{hunter_Matplotlib2DGraphics_2007}
Hunter, J.~D. 2007, \bibinfo{title}{Matplotlib: {A} {2D} graphics environment,} Computing in Science \& Engineering, 9, 90, \dodoi{10.1109/MCSE.2007.55}

\bibitem[{R.~E. Hviding(2025)Hviding}]{hviding_TheSkyentistUniteVersion_2025}
Hviding, R.~E. 2025, {TheSkyentist}/unite: {Version} 0, \dodoi{10.5281/zenodo.15585035}

\bibitem[{R.~E. Hviding {et~al.}(2025)Hviding, de~Graaff, Miller, Setton, Greene, Labbé, Brammer, Bezanson, Boogaard, Cleri, Leja, Maseda, McConachie, Matthee, Naidu, Oesch, Wang, Whitaker, \& Williams}]{hviding_RUBIESSpectroscopicCensus_2025}
Hviding, R.~E., de~Graaff, A., Miller, T.~B., {et~al.} 2025, \bibinfo{title}{{RUBIES}: {A} spectroscopic census of little red dots: {All} point sources with v-{Shaped} continua have broad lines,} Astronomy and Astrophysics, 702, A57, \dodoi{10.1051/0004-6361/202555816}

\bibitem[{K. Inayoshi(2025)Inayoshi}]{inayoshi_LittleRedDots_2025}
Inayoshi, K. 2025, \bibinfo{title}{Little {Red} {Dots} as the {Very} {First} {Activity} of {Black} {Hole} {Growth},} The Astrophysical Journal, 988, L22, \dodoi{10.3847/2041-8213/adea66}

\bibitem[{K. Inayoshi \& R. Maiolino(2025)Inayoshi \& Maiolino}]{inayoshi_ExtremelyDenseGas_2025}
Inayoshi, K., \& Maiolino, R. 2025, \bibinfo{title}{Extremely {Dense} {Gas} around {Little} {Red} {Dots} and {High}-redshift {Active} {Galactic} {Nuclei}: {A} {Nonstellar} {Origin} of the {Balmer} {Break} and {Absorption} {Features},} The Astrophysical Journal, 980, L27, \dodoi{10.3847/2041-8213/adaebd}

\bibitem[{X. Ji {et~al.}(2025)Ji, Maiolino, Übler, Scholtz, D'Eugenio, Sun, Perna, Turner, Arribas, Bennett, Bunker, Carniani, Charlot, Cresci, Curti, Egami, Fabian, Inayoshi, Isobe, Jones, Juodžbalis, Kumari, Lyu, Mazzolari, Parlanti, Robertson, Rodrı́guez Del~Pino, Schneider, Sijacki, Tacchella, Trinca, Valiante, Venturi, Volonteri, Willott, Witten, \& Witstok}]{ji_BlackTHUNDERNonstellarBalmer_2025}
Ji, X., Maiolino, R., Übler, H., {et~al.} 2025, \bibinfo{title}{{BlackTHUNDER} – {A} non-stellar {Balmer} break in a black hole-dominated little red dot at {\textless}span class="nocase"{\textgreater}z=7.04{\textless}/span{\textgreater},} arXiv, arXiv:2501.13082, \dodoi{10.48550/arXiv.2501.13082}

\bibitem[{C. Jin {et~al.}(2012)Jin, Ward, \& Done}]{jin_CombinedOpticalXray_2012}
Jin, C., Ward, M., \& Done, C. 2012, \bibinfo{title}{A combined optical and {X}-ray study of unobscured type 1 active galactic nuclei - {II}. {Relation} between {X}-ray emission and optical spectra,} Monthly Notices of the Royal Astronomical Society, 422, 3268, \dodoi{10.1111/j.1365-2966.2012.20847.x}

\bibitem[{B. Johnson \& J. Leja(2017)Johnson \& Leja}]{johnson_BdjProspectorInitial_2017}
Johnson, B., \& Leja, J. 2017, Bd-j/prospector: {Initial} release, Zenodo, \dodoi{10.5281/zenodo.1116491}

\bibitem[{A.~P. Jones {et~al.}(2017)Jones, Köhler, Ysard, Bocchio, \& Verstraete}]{jones_GlobalDustModelling_2017}
Jones, A.~P., Köhler, M., Ysard, N., Bocchio, M., \& Verstraete, L. 2017, \bibinfo{title}{The global dust modelling framework {THEMIS},} Astronomy and Astrophysics, 602, A46, \dodoi{10.1051/0004-6361/201630225}

\bibitem[{I. Juodžbalis {et~al.}(2024)Juodžbalis, Ji, Maiolino, D'Eugenio, Scholtz, Risaliti, Fabian, Mazzolari, Gilli, Prandoni, Arribas, Bunker, Carniani, Charlot, Curtis-Lake, de~Graaff, Hainline, Parlanti, Perna, Pérez-González, Robertson, Tacchella, Übler, Williams, Willott, \& Witstok}]{juodzbalis_JADESRosettaStone_2024}
Juodžbalis, I., Ji, X., Maiolino, R., {et~al.} 2024, \bibinfo{title}{{JADES} - the {Rosetta} stone of {JWST}-discovered {AGN}: {Deciphering} the intriguing nature of early {AGN},} Monthly Notices of the Royal Astronomical Society, 535, 853, \dodoi{10.1093/mnras/stae2367}

\bibitem[{D. Kido {et~al.}(2025)Kido, Ioka, Hotokezaka, Inayoshi, \& Irwin}]{kido_BlackHoleEnvelopes_2025}
Kido, D., Ioka, K., Hotokezaka, K., Inayoshi, K., \& Irwin, C.~M. 2025, \bibinfo{title}{Black hole envelopes in little red dots,} arXiv, arXiv:2505.06965, \dodoi{10.48550/arXiv.2505.06965}

\bibitem[{D.~D. Kocevski {et~al.}(2025)Kocevski, Finkelstein, Barro, Taylor, Calabrò, Laloux, Buchner, Trump, Leung, Yang, Dickinson, Pérez-González, Pacucci, Inayoshi, Somerville, McGrath, Akins, Bagley, Bowler, Bisigello, Carnall, Casey, Cheng, Cleri, Costantin, Cullen, Davis, Donnan, Dunlop, Ellis, Ferguson, Fujimoto, Fontana, Giavalisco, Grazian, Grogin, Hathi, Hirschmann, Huertas-Company, Holwerda, Illingworth, Juneau, Kartaltepe, Koekemoer, Li, Lucas, Magee, Mason, McLeod, McLure, Napolitano, Papovich, Pirzkal, Rodighiero, Santini, Wilkins, \& Yung}]{kocevski_RiseFaintRed_2025}
Kocevski, D.~D., Finkelstein, S.~L., Barro, G., {et~al.} 2025, \bibinfo{title}{The {Rise} of {Faint}, {Red} {Active} {Galactic} {Nuclei} at z {\textgreater} 4: {A} {Sample} of {Little} {Red} {Dots} in the {JWST} {Extragalactic} {Legacy} {Fields},} The Astrophysical Journal Letters, 986, 126, \dodoi{10.3847/1538-4357/adbc7d}

\bibitem[{V. Kokorev {et~al.}(2024)Kokorev, Caputi, Greene, Dayal, Trebitsch, Cutler, Fujimoto, Labbé, Miller, Iani, Navarro-Carrera, \& Rinaldi}]{kokorev_CensusPhotometricallySelected_2024}
Kokorev, V., Caputi, K.~I., Greene, J.~E., {et~al.} 2024, \bibinfo{title}{A {Census} of {Photometrically} {Selected} {Little} {Red} {Dots} at 4 {\textless} z {\textless} 9 in {JWST} {Blank} {Fields},} The Astrophysical Journal, 968, 38, \dodoi{10.3847/1538-4357/ad4265}

\bibitem[{K.~T. Korista \& M.~R. Goad(2004)Korista \& Goad}]{korista_WhatOpticalRecombination_2004}
Korista, K.~T., \& Goad, M.~R. 2004, \bibinfo{title}{What the {Optical} {Recombination} {Lines} {Can} {Tell} {Us} about the {Broad}-{Line} {Regions} of {Active} {Galactic} {Nuclei},} The Astrophysical Journal, 606, 749, \dodoi{10.1086/383193}

\bibitem[{I. Labbe {et~al.}(2024)Labbe, Greene, Matthee, Treiber, Kokorev, Miller, Kramarenko, Setton, Ma, Goulding, Bezanson, Naidu, Williams, Atek, Brammer, Cutler, Chemerynska, Cloonan, Dayal, de~Graaff, Fudamoto, Fujimoto, Furtak, Glazebrook, Heintz, Leja, Marchesini, Nanayakkara, Nelson, Oesch, Pan, Price, Shivaei, Sobral, Suess, van Dokkum, Wang, Weaver, Whitaker, \& Zitrin}]{labbe_UnambiguousAGNBalmer_2024}
Labbe, I., Greene, J.~E., Matthee, J., {et~al.} 2024, \bibinfo{title}{An unambiguous {AGN} and a balmer break in an ultraluminous little red dot at {Z}=4.47 from ultradeep {UNCOVER} and all the little things spectroscopy,} arXiv, arXiv:2412.04557, \dodoi{10.48550/arXiv.2412.04557}

\bibitem[{E. Lambrides {et~al.}(2024)Lambrides, Garofali, Larson, Ptak, Chiaberge, Long, Hutchison, Norman, McKinney, Akins, Berg, Chisholm, Civano, Cloonan, Endsley, Faisst, Gilli, Gillman, Hirschmann, Kartaltepe, Kocevski, Kokorev, Pacucci, Richardson, Stiavelli, \& Whalen}]{lambrides_CaseSupereddingtonAccretion_2024}
Lambrides, E., Garofali, K., Larson, R., {et~al.} 2024, \bibinfo{title}{The case for super-eddington accretion: {Connecting} weak x-{Ray} and {UV} line emission in {JWST} broad-line {AGN} during the first gyr of cosmic time,} arXiv e-prints, arXiv:2409.13047, \dodoi{10.48550/arXiv.2409.13047}

\bibitem[{L. Lamport(1994)Lamport}]{lamport_LaTeXDocumentPreparation_1994}
Lamport, L. 1994, {LaTeX}: {A} {Document} {Preparation} {System}, 2nd edn. (Addison-Wesley Professional)

\bibitem[{H. Liu {et~al.}(2025)Liu, Jiang, Quataert, Greene, \& Ma}]{liu_BalmerBreakOptical_2025}
Liu, H., Jiang, Y.-F., Quataert, E., Greene, J.~E., \& Ma, Y. 2025, \bibinfo{title}{The {Balmer} {Break} and {Optical} {Continuum} of {Little} {Red} {Dots} from {Super}-{Eddington} {Accretion},} The Astrophysical Journal, 994, 113, \dodoi{10.3847/1538-4357/ae0c19}

\bibitem[{I.~E. López {et~al.}(2024)López, Yang, Mountrichas, Brusa, Alexander, Baldi, Bertola, Bonoli, Comastri, Shankar, Acharya, Alonso~Tetilla, Lapi, Laloux, López~López, Muñoz~Rodríguez, Musiimenta, Osorio~Clavijo, Sala, \& Sengupta}]{lopez_CIGALEModuleTailored_2024}
López, I.~E., Yang, G., Mountrichas, G., {et~al.} 2024, \bibinfo{title}{A {CIGALE} module tailored (not only) for low-luminosity active galactic nuclei,} Astronomy and Astrophysics, 692, A209, \dodoi{10.1051/0004-6361/202450510}

\bibitem[{Y. Ma {et~al.}(2025{\natexlab{a}})Ma, Greene, Setton, Goulding, Annunziatella, Fan, Kokorev, Labbe, Li, Lin, Marchesini, Matthee, Robbins, Sajina, Sawicki, \& Telford}]{ma_CountingLittleRed_2025}
Ma, Y., Greene, J.~E., Setton, D.~J., {et~al.} 2025{\natexlab{a}}, \bibinfo{title}{Counting {Little} {Red} {Dots} at \$z{\textless}4\$ with {Ground}-based {Surveys} and {Spectroscopic} {Follow}-up,} arXiv, arXiv:2504.08032, \dodoi{10.48550/arXiv.2504.08032}

\bibitem[{Y. Ma {et~al.}(2025{\natexlab{b}})Ma, Greene, Setton, Volonteri, Leja, Wang, Bezanson, Brammer, Cutler, Dayal, van Dokkum, Furtak, Glazebrook, Goulding, de~Graaff, Kokorev, Labbe, Pan, Price, Weaver, Williams, Whitaker, \& Zitrin}]{ma_UNCOVER404Error_2025}
Ma, Y., Greene, J.~E., Setton, D.~J., {et~al.} 2025{\natexlab{b}}, \bibinfo{title}{{UNCOVER}: 404 {Error}—{Models} {Not} {Found} for the {Triply} {Imaged} {Little} {Red} {Dot} {A2744}-{QSO1},} The Astrophysical Journal, 981, 191, \dodoi{10.3847/1538-4357/ada613}

\bibitem[{P. Madau \& F. Haardt(2024)Madau \& Haardt}]{madau_XRayWeakActive_2024}
Madau, P., \& Haardt, F. 2024, \bibinfo{title}{X-{Ray} {Weak} {Active} {Galactic} {Nuclei} from {Super}-{Eddington} {Accretion} onto {Infant} {Black} {Holes},} The Astrophysical Journal, 976, L24, \dodoi{10.3847/2041-8213/ad90e1}

\bibitem[{R. Maiolino {et~al.}(2001)Maiolino, Marconi, Salvati, Risaliti, Severgnini, Oliva, La~Franca, \& Vanzi}]{maiolino_DustActiveNuclei_2001}
Maiolino, R., Marconi, A., Salvati, M., {et~al.} 2001, \bibinfo{title}{Dust in active nuclei. {I}. {Evidence} for “anomalous” properties,} Astronomy and Astrophysics, 365, 28, \dodoi{10.1051/0004-6361:20000177}

\bibitem[{R. Maiolino {et~al.}(2025)Maiolino, Risaliti, Signorini, Trefoloni, Juodžbalis, Scholtz, Übler, D'Eugenio, Carniani, Fabian, Ji, Mazzolari, Bertola, Brusa, Bunker, Charlot, Comastri, Cresci, DeCoursey, Egami, Fiore, Gilli, Perna, Tacchella, \& Venturi}]{maiolino_JWSTMeetsChandra_2025}
Maiolino, R., Risaliti, G., Signorini, M., {et~al.} 2025, \bibinfo{title}{{JWST} meets {Chandra}: a large population of {Compton} thick, feedback-free, and intrinsically {X}-ray weak {AGN}, with a sprinkle of {SNe},} Monthly Notices of the Royal Astronomical Society, 538, 1921, \dodoi{10.1093/mnras/staf359}

\bibitem[{L.~N. Martínez-Ramírez {et~al.}(2024)Martínez-Ramírez, Calistro~Rivera, Lusso, Bauer, Nardini, Buchner, Brown, Pineda, Temple, Banerji, Stalevski, \& Hennawi}]{martinez-ramirez_AGNFITTERRXModelingRadiotoXray_2024}
Martínez-Ramírez, L.~N., Calistro~Rivera, G., Lusso, E., {et~al.} 2024, \bibinfo{title}{{AGNFITTER}-{RX}: {Modeling} the radio-to-{X}-ray spectral energy distributions of {AGNs},} Astronomy and Astrophysics, 688, A46, \dodoi{10.1051/0004-6361/202449329}

\bibitem[{M.~V. Maseda {et~al.}(2024)Maseda, de~Graaff, Franx, Rix, Carniani, Laseter, Dudzevičiūtė, Rawle, Parlanti, Arribas, Bunker, Cameron, Charlot, Curti, D'Eugenio, Jones, Kumari, Maiolino, Übler, Saxena, Smit, Willott, \& Witstok}]{maseda_NIRSpecWideGTO_2024}
Maseda, M.~V., de~Graaff, A., Franx, M., {et~al.} 2024, \bibinfo{title}{The {NIRSpec} {Wide} {GTO} {Survey},} Astronomy and Astrophysics, 689, A73, \dodoi{10.1051/0004-6361/202449914}

\bibitem[{J. Matthee {et~al.}(2024)Matthee, Naidu, Brammer, Chisholm, Eilers, Goulding, Greene, Kashino, Labbe, Lilly, Mackenzie, Oesch, Weibel, Wuyts, Xiao, Bordoloi, Bouwens, van Dokkum, Illingworth, Kramarenko, Maseda, Mason, Meyer, Nelson, Reddy, Shivaei, Simcoe, \& Yue}]{matthee_LittleRedDots_2024}
Matthee, J., Naidu, R.~P., Brammer, G., {et~al.} 2024, \bibinfo{title}{Little {Red} {Dots}: {An} {Abundant} {Population} of {Faint} {Active} {Galactic} {Nuclei} at z ∼ 5 {Revealed} by the {EIGER} and {FRESCO} {JWST} {Surveys},} The Astrophysical Journal, 963, 129, \dodoi{10.3847/1538-4357/ad2345}

\bibitem[{R.~P. Naidu {et~al.}(2025)Naidu, Matthee, Katz, de~Graaff, Oesch, Smith, Greene, Brammer, Weibel, Hviding, Chisholm, Labb{\textbackslash}'e, Simcoe, Witten, Atek, Baggen, Belli, Bezanson, Boogaard, Bose, Covelo-Paz, Dayal, Fudamoto, Furtak, Giovinazzo, Goulding, Gronke, Heintz, Hirschmann, Illingworth, Inoue, Johnson, Leja, Leonova, McConachie, Maseda, Natarajan, Nelson, Setton, Shivaei, Sobral, Stefanon, Tacchella, Toft, Torralba, van Dokkum, van~der Wel, Volonteri, Walter, Wang, \& Watson}]{naidu_BlackHoleStar_2025}
Naidu, R.~P., Matthee, J., Katz, H., {et~al.} 2025, \bibinfo{title}{A “black hole star” reveals the remarkable gas-enshrouded hearts of the little red dots,} arXiv, arXiv:2503.16596, \dodoi{10.48550/arXiv.2503.16596}

\bibitem[{S. Noll {et~al.}(2009)Noll, Burgarella, Giovannoli, Buat, Marcillac, \& Muñoz-Mateos}]{noll_AnalysisGalaxySpectral_2009}
Noll, S., Burgarella, D., Giovannoli, E., {et~al.} 2009, \bibinfo{title}{Analysis of galaxy spectral energy distributions from far-{UV} to far-{IR} with {CIGALE}: {Studying} a {SINGS} test sample,} Astronomy and Astrophysics, 507, 1793, \dodoi{10.1051/0004-6361/200912497}

\bibitem[{T.~E. Oliphant(2006)Oliphant}]{oliphant_GuideNumPy_2006}
Oliphant, T.~E. 2006, A guide to {NumPy}, Vol.~1 (Trelgol Publishing USA)

\bibitem[{D. Phan {et~al.}(2019)Phan, Pradhan, \& Jankowiak}]{phan_ComposableEffectsFlexible_2019}
Phan, D., Pradhan, N., \& Jankowiak, M. 2019, \bibinfo{title}{Composable effects for flexible and accelerated probabilistic programming in {NumPyro},} arXiv e-prints, arXiv:1912.11554

\bibitem[{P. Rinaldi {et~al.}(2024)Rinaldi, Bonaventura, Rieke, Alberts, Caputi, Baker, Baum, Bhatawdekar, Bunker, Carniani, Curtis-Lake, D'Eugenio, Egami, Ji, Hainline, Helton, Lin, Lyu, Johnson, Ma, Maiolino, Pérez-González, Rieke, Robertson, Shivaei, Stone, Sun, Tacchella, Übler, Williams, Willmer, Willott, Zhang, \& Zhu}]{rinaldi_NotJustDot_2024}
Rinaldi, P., Bonaventura, N., Rieke, G.~H., {et~al.} 2024, \bibinfo{title}{Not just a dot: {The} complex {UV} morphology and underlying properties of little red dots,} arXiv e-prints, arXiv:2411.14383, \dodoi{10.48550/arXiv.2411.14383}

\bibitem[{V. Rusakov {et~al.}(2025)Rusakov, Watson, Nikopoulos, Brammer, Gottumukkala, Harvey, Heintz, Nielsen, Sim, Sneppen, Vijayan, Adams, Austin, Conselice, Goolsby, Toft, \& Witstok}]{rusakov_JWSTsLittleRed_2025}
Rusakov, V., Watson, D., Nikopoulos, G.~P., {et~al.} 2025, \bibinfo{title}{{JWST}'s little red dots: {An} emerging population of young, low-mass {AGN} cocooned in dense ionized gas,} arXiv, arXiv:2503.16595, \dodoi{10.48550/arXiv.2503.16595}

\bibitem[{S. Salim {et~al.}(2018)Salim, Boquien, \& Lee}]{salim_DustAttenuationCurves_2018}
Salim, S., Boquien, M., \& Lee, J.~C. 2018, \bibinfo{title}{Dust {Attenuation} {Curves} in the {Local} {Universe}: {Demographics} and {New} {Laws} for {Star}-forming {Galaxies} and {High}-redshift {Analogs},} The Astrophysical Journal, 859, 11, \dodoi{10.3847/1538-4357/aabf3c}

\bibitem[{A.~D. Santarelli {et~al.}(2025)Santarelli, Farag, Bellinger, Natarajan, Naidu, Campbell, \& Caplan}]{santarelli_EvolutionaryTracksSpectral_2025}
Santarelli, A.~D., Farag, E., Bellinger, E.~P., {et~al.} 2025, Evolutionary {Tracks} and {Spectral} {Properties} of {Quasi}-stars and {Their} {Correlation} with {Little} {Red} {Dots}, arXiv, \dodoi{10.48550/arXiv.2510.17952}

\bibitem[{C. Schreiber {et~al.}(2017)Schreiber, Elbaz, Pannella, Merlin, Castellano, Fontana, Bourne, Boutsia, Cullen, Dunlop, Ferguson, Michałowski, Okumura, Santini, Shu, Wang, \& White}]{schreiber_EGGHatchingMock_2017}
Schreiber, C., Elbaz, D., Pannella, M., {et~al.} 2017, \bibinfo{title}{{EGG}: {Hatching} a mock {Universe} from empirical prescriptions⋆,} Astronomy and Astrophysics, 602, A96, \dodoi{10.1051/0004-6361/201629123}

\bibitem[{D.~J. Setton {et~al.}(2024)Setton, Greene, de~Graaff, Ma, Leja, Matthee, Bezanson, Boogaard, Cleri, Katz, Labbe, Maseda, McConachie, Miller, Price, Suess, van Dokkum, Wang, Weibel, Whitaker, \& Williams}]{setton_LittleRedDots_2024}
Setton, D.~J., Greene, J.~E., de~Graaff, A., {et~al.} 2024, \bibinfo{title}{Little red dots at an inflection point: {Ubiquitous} “v-{Shaped}” turnover consistently occurs at the balmer limit,} arXiv, arXiv:2411.03424, \dodoi{10.48550/arXiv.2411.03424}

\bibitem[{D.~J. Setton {et~al.}(2025)Setton, Greene, Spilker, Williams, Labbe, Ma, Wang, Whitaker, Leja, de~Graaff, Alberts, Bezanson, Boogaard, Brammer, Cutler, Cleri, Cooper, Dayal, Fujimoto, Furtak, Goulding, Hirschmann, Kokorev, Maseda, McConachie, Matthee, Miller, Naidu, Oesch, Pan, Price, Suess, Weaver, Xiao, Zhang, \& Zitrin}]{setton_ConfirmedDeficitHot_2025}
Setton, D.~J., Greene, J.~E., Spilker, J.~S., {et~al.} 2025, \bibinfo{title}{A confirmed deficit of hot and cold dust emission in the most luminous {Little} {Red} {Dots},} arXiv e-prints, arXiv:2503.02059, \dodoi{10.48550/arXiv.2503.02059}

\bibitem[{R.~E. Skelton {et~al.}(2014)Skelton, Whitaker, Momcheva, Brammer, van Dokkum, Labbé, Franx, van~der Wel, Bezanson, Da~Cunha, Fumagalli, Förster~Schreiber, Kriek, Leja, Lundgren, Magee, Marchesini, Maseda, Nelson, Oesch, Pacifici, Patel, Price, Rix, Tal, Wake, \& Wuyts}]{skelton_3DHSTWFC3selectedPhotometric_2014}
Skelton, R.~E., Whitaker, K.~E., Momcheva, I.~G., {et~al.} 2014, \bibinfo{title}{{3D}-{HST} {WFC3}-selected {Photometric} {Catalogs} in the {Five} {CANDELS}/{3D}-{HST} {Fields}: {Photometry}, {Photometric} {Redshifts}, and {Stellar} {Masses},} The Astrophysical Journal Supplement Series, 214, 24, \dodoi{10.1088/0067-0049/214/2/24}

\bibitem[{M. Stalevski {et~al.}(2012)Stalevski, Fritz, Baes, Nakos, \& Popović}]{stalevski_3DRadiativeTransfer_2012}
Stalevski, M., Fritz, J., Baes, M., Nakos, T., \& Popović, L.~{\v C}. 2012, \bibinfo{title}{{3D} radiative transfer modelling of the dusty tori around active galactic nuclei as a clumpy two-phase medium,} Monthly Notices of the Royal Astronomical Society, 420, 2756, \dodoi{10.1111/j.1365-2966.2011.19775.x}

\bibitem[{M. Stalevski {et~al.}(2016)Stalevski, Ricci, Ueda, Lira, Fritz, \& Baes}]{stalevski_DustCoveringFactor_2016}
Stalevski, M., Ricci, C., Ueda, Y., {et~al.} 2016, \bibinfo{title}{The dust covering factor in active galactic nuclei,} Monthly Notices of the Royal Astronomical Society, 458, 2288, \dodoi{10.1093/mnras/stw444}

\bibitem[{D. Stern(2015)Stern}]{stern_XRayMidinfraredRelation_2015}
Stern, D. 2015, \bibinfo{title}{The {X}-{Ray} to {Mid}-infrared {Relation} of {AGNs} at {High} {Luminosity},} The Astrophysical Journal, 807, 129, \dodoi{10.1088/0004-637X/807/2/129}

\bibitem[{A.~J. Taylor {et~al.}(2025)Taylor, Kokorev, Kocevski, Akins, Cullen, Dickinson, Finkelstein, Arrabal~Haro, Bromm, Giavalisco, Inayoshi, Juneau, Leung, Perez-Gonzalez, Somerville, Trump, Amorin, Barro, Burgarella, Brooks, Carnall, Casey, Cheng, Chisholm, Chworowsky, Davis, Donnan, Dunlop, Ellis, Fernandez, Fujimoto, Grogin, Gupta, Hathi, Jung, Hirschmann, Kartaltepe, Koekemoer, Larson, Leung, Llerena, Lucas, McLeod, McLure, Napolitano, Papovich, Stanton, Tripodi, Wang, Wilkins, Yung, \& Zavala}]{taylor_CAPERSLRDz9GasEnshrouded_2025}
Taylor, A.~J., Kokorev, V., Kocevski, D.~D., {et~al.} 2025, \bibinfo{title}{{CAPERS}-{LRD}-z9: a gas enshrouded little red dot hosting a broad-line {AGN} at {Z}=9.288,} arXiv, arXiv:2505.04609, \dodoi{10.48550/arXiv.2505.04609}

\bibitem[{M.~J. Temple {et~al.}(2021)Temple, Hewett, \& Banerji}]{temple_ModellingType1_2021}
Temple, M.~J., Hewett, P.~C., \& Banerji, M. 2021, \bibinfo{title}{Modelling type 1 quasar colours in the era of {Rubin} and {Euclid},} Monthly Notices of the Royal Astronomical Society, 508, 737, \dodoi{10.1093/mnras/stab2586}

\bibitem[{A. Torralba {et~al.}(2025)Torralba, Matthee, Pezzulli, Naidu, Ishikawa, Brammer, Chang, Chisholm, de~Graaff, D'Eugenio, Cesare, Eilers, Greene, Gronke, Iani, Kokorev, Kotiwale, Kramarenko, Ma, Mascia, Navarrete, Nelson, Oesch, Simcoe, \& Wuyts}]{torralba_WarmOuterLayer_2025}
Torralba, A., Matthee, J., Pezzulli, G., {et~al.} 2025, \bibinfo{title}{The warm outer layer of a {Little} {Red} {Dot} as the source of [{Fe} {II}] and collisional {Balmer} lines with scattering wings,} arXiv, arXiv:2510.00103, \dodoi{10.48550/arXiv.2510.00103}

\bibitem[{A. Tortosa {et~al.}(2023)Tortosa, Ricci, Ho, Tombesi, Du, Inayoshi, Wang, Shangguan, \& Li}]{tortosa_SystematicBroadbandXray_2023}
Tortosa, A., Ricci, C., Ho, L.~C., {et~al.} 2023, \bibinfo{title}{Systematic broad-band {X}-ray study of super-{Eddington} accretion on to supermassive black holes - {I}. {X}-ray continuum,} Monthly Notices of the Royal Astronomical Society, 519, 6267, \dodoi{10.1093/mnras/stac3590}

\bibitem[{H. Umeda {et~al.}(2025)Umeda, Inayoshi, Harikane, \& Murase}]{umeda_BlackHoleEnvelopeInterpretation_2025}
Umeda, H., Inayoshi, K., Harikane, Y., \& Murase, K. 2025, \bibinfo{title}{A {Black}-{Hole} {Envelope} {Interpretation} for {Cosmological} {Demographics} of {Little} {Red} {Dots},} arXiv e-prints, arXiv:2512.04208, \dodoi{10.48550/arXiv.2512.04208}

\bibitem[{S. van~der Walt {et~al.}(2011)van~der Walt, Colbert, \& Varoquaux}]{vanderwalt_NumPyArrayStructure_2011}
van~der Walt, S., Colbert, S.~C., \& Varoquaux, G. 2011, \bibinfo{title}{The {NumPy} {Array}: {A} {Structure} for {Efficient} {Numerical} {Computation},} Computing in Science Engineering, 13, 22, \dodoi{10.1109/MCSE.2011.37}

\bibitem[{A. Van Der~Wel {et~al.}(2012)Van Der~Wel, Bell, Häussler, McGrath, Chang, Guo, McIntosh, Rix, Barden, Cheung, Faber, Ferguson, Galametz, Grogin, Hartley, Kartaltepe, Kocevski, Koekemoer, Lotz, Mozena, Peth, \& Peng}]{vanderwel_StructuralParametersGalaxies_2012}
Van Der~Wel, A., Bell, E.~F., Häussler, B., {et~al.} 2012, \bibinfo{title}{Structural parameters of galaxies in candels,} Astrophysical Journal, Supplement Series, 203, \dodoi{10.1088/0067-0049/203/2/24}

\bibitem[{D.~E. Vanden~Berk {et~al.}(2001)Vanden~Berk, Richards, Bauer, Strauss, Schneider, Heckman, York, Hall, Fan, Knapp, Anderson, Annis, Bahcall, Bernardi, Briggs, Brinkmann, Brunner, Burles, Carey, Castander, Connolly, Crocker, Csabai, Doi, Finkbeiner, Friedman, Frieman, Fukugita, Gunn, Hennessy, Ivezić, Kent, Kunszt, Lamb, Leger, Long, Loveday, Lupton, Meiksin, Merelli, Munn, Newberg, Newcomb, Nichol, Owen, Pier, Pope, Rockosi, Schlegel, Siegmund, Smee, Snir, Stoughton, Stubbs, SubbaRao, Szalay, Szokoly, Tremonti, Uomoto, Waddell, Yanny, \& Zheng}]{vandenberk_CompositeQuasarSpectra_2001}
Vanden~Berk, D.~E., Richards, G.~T., Bauer, A., {et~al.} 2001, \bibinfo{title}{Composite {Quasar} {Spectra} from the {Sloan} {Digital} {Sky} {Survey},} The Astronomical Journal, 122, 549, \dodoi{10.1086/321167}

\bibitem[{P. Virtanen {et~al.}(2020)Virtanen, Gommers, Oliphant, Haberland, Reddy, Cournapeau, Burovski, Peterson, Weckesser, Bright, van~der Walt, Brett, Wilson, Jarrod~Millman, Mayorov, Nelson, Jones, Kern, Larson, Carey, Polat, Feng, Moore, Vand~erPlas, Laxalde, Perktold, Cimrman, Henriksen, Quintero, Harris, Archibald, Ribeiro, Pedregosa, van Mulbregt, \& Contributors}]{virtanen_SciPy10Fundamental_2020}
Virtanen, P., Gommers, R., Oliphant, T.~E., {et~al.} 2020, \bibinfo{title}{{SciPy} 1.0: {Fundamental} {Algorithms} for {Scientific} {Computing} in {Python},} Nature Methods, 17, 261, \dodoi{10.1038/s41592-019-0686-2}

\bibitem[{B. Wang {et~al.}(2024)Wang, Leja, de~Graaff, Brammer, Weibel, van Dokkum, Baggen, Suess, Greene, Bezanson, Cleri, Hirschmann, Labbé, Matthee, McConachie, Naidu, Nelson, Oesch, Setton, \& Williams}]{wang_RUBIESEvolvedStellar_2024}
Wang, B., Leja, J., de~Graaff, A., {et~al.} 2024, \bibinfo{title}{{RUBIES}: {Evolved} {Stellar} {Populations} with {Extended} {Formation} {Histories} at z ∼ 7–8 in {Candidate} {Massive} {Galaxies} {Identified} with {JWST}/{NIRSpec},} The Astrophysical Journal, 969, L13, \dodoi{10.3847/2041-8213/ad55f7}

\bibitem[{B. Wang {et~al.}(2025)Wang, de~Graaff, Davies, Greene, Leja, Brammer, Goulding, Miller, Suess, Weibel, Williams, Bezanson, Boogaard, Cleri, Hirschmann, Katz, Labbé, Maseda, Matthee, McConachie, Naidu, Oesch, Rix, Setton, \& Whitaker}]{wang_RUBIESJWSTNIRSpec_2025}
Wang, B., de~Graaff, A., Davies, R.~L., {et~al.} 2025, \bibinfo{title}{{RUBIES}: {JWST}/{NIRSpec} {Confirmation} of an {Infrared}-luminous, {Broad}-line {Little} {Red} {Dot} with an {Ionized} {Outflow},} The Astrophysical Journal, 984, 121, \dodoi{10.3847/1538-4357/adc1ca}

\bibitem[{S. Watanabe \& M. Opper(2010)Watanabe \& Opper}]{watanabe_AsymptoticEquivalenceBayes_2010}
Watanabe, S., \& Opper, M. 2010, \bibinfo{title}{Asymptotic equivalence of {Bayes} cross validation and widely applicable information criterion in singular learning theory.,} Journal of machine learning research, 11

\bibitem[{C.~C. Williams {et~al.}(2024)Williams, Alberts, Ji, Hainline, Lyu, Rieke, Endsley, Suess, Sun, Johnson, Florian, Shivaei, Rujopakarn, Baker, Bhatawdekar, Boyett, Bunker, Cameron, Carniani, Charlot, Curtis-Lake, DeCoursey, de~Graaff, Egami, Eisenstein, Gibson, Hausen, Helton, Maiolino, Maseda, Nelson, Pérez-González, Rieke, Robertson, Saxena, Tacchella, Willmer, \& Willott}]{williams_GalaxiesMissedHubble_2024}
Williams, C.~C., Alberts, S., Ji, Z., {et~al.} 2024, \bibinfo{title}{The {Galaxies} {Missed} by {Hubble} and {ALMA}: {The} {Contribution} of {Extremely} {Red} {Galaxies} to the {Cosmic} {Census} at 3 {\textless} z {\textless} 8,} The Astrophysical Journal, 968, 34, \dodoi{10.3847/1538-4357/ad3f17}

\bibitem[{G. Yang {et~al.}(2020)Yang, Boquien, Buat, Burgarella, Ciesla, Duras, Stalevski, Brandt, \& Papovich}]{yang_XCIGALEFittingAGN_2020}
Yang, G., Boquien, M., Buat, V., {et~al.} 2020, \bibinfo{title}{X-{CIGALE}: {Fitting} {AGN}/galaxy {SEDs} from {X}-ray to infrared,} Monthly Notices of the Royal Astronomical Society, 491, 740, \dodoi{10.1093/mnras/stz3001}

\bibitem[{M. Yue {et~al.}(2024)Yue, Eilers, Ananna, Panagiotou, Kara, \& Miyaji}]{yue_StackingXRayObservations_2024}
Yue, M., Eilers, A.-C., Ananna, T.~T., {et~al.} 2024, \bibinfo{title}{Stacking {X}-{Ray} {Observations} of "{Little} {Red} {Dots}": {Implications} for {Their} {Active} {Galactic} {Nucleus} {Properties},} The Astrophysical Journal, 974, L26, \dodoi{10.3847/2041-8213/ad7eba}

\end{thebibliography}
\bibliographystyle{aasjournalv7}

\end{document}